\documentclass[fleqn,usenatbib]{mnras}

\usepackage{newtxtext,newtxmath}

\usepackage[T1]{fontenc}

\DeclareRobustCommand{\VAN}[3]{#2}
\let\VANthebibliography\thebibliography
\def\thebibliography{\DeclareRobustCommand{\VAN}[3]{##3}\VANthebibliography}

\usepackage{graphicx}	
\usepackage{amsmath}	

\usepackage{amssymb}	

\usepackage{multirow}

\newcommand{\HII}{\rm H~{\sc ii }}

\newcommand{\XHIIv}{$x^{\rm v}_{\rm HII}$}

\newcommand{\bettiZero}{$\beta_0~$}
\newcommand{\bettiOne}{$\beta_1~$}
\newcommand{\bettiTwo}{$\beta_2~$}
\newcommand{\CommentOut}[1]{}
\newcommand{\RefReport}[1]{{#1}}

\title[Topology of reionization with Betti numbers]{Measuring the topology of reionization with Betti numbers}

\author[Giri \& Mellema \CommentOut{et al.}]{
Sambit K. Giri,$^{1,2}$\thanks{E-mail: sambit.giri@gmail.com} and Garrelt Mellema$^{2}$
\\
$^{1}$Institute for Computational Science, University of Zurich, Winterthurerstrasse 190, 8057
Zurich, Switzerland \\
$^{2}$Department of Astronomy and Oskar Klein Center, Stockholm University, AlbaNova, SE-106 91 Stockholm, Sweden\\
}

\date{Accepted XXX. Received YYY; in original form ZZZ}

\pubyear{2020}

\begin{document}
\label{firstpage}
\pagerange{\pageref{firstpage}--\pageref{lastpage}}
\maketitle

\begin{abstract}
The distribution of ionised hydrogen during the epoch of reionization (EoR) has a complex morphology. We propose to measure the three-dimensional topology of ionised  regions using the Betti numbers. These quantify the topology using the number of components, tunnels and cavities in any given field. Based on the results for a set of reionization simulations we find that the Betti numbers of the ionisation field show a characteristic evolution during reionization, with peaks in the different Betti numbers characterising different stages of the process. The shapes of their evolutionary curves can be fitted with simple analytical functions. We also observe that the evolution of the Betti numbers shows a clear connection with the percolation of the ionized and neutral regions and differs between different reionization scenarios. Through these properties, the Betti numbers provide a more useful description of the topology than the widely studied Euler characteristic or genus. The morphology of the ionisation field will be imprinted on the redshifted 21-cm signal from the EoR. We construct mock image cubes using the properties of the low-frequency element of the future Square Kilometre Array and show that we can extract the Betti numbers from such data sets if an observation time of 1000 h is used. Even for a much shorter observation time of 100 h, some topological information can be extracted for the middle and later stages of reionization. We also find that the topological information extracted from the mock 21-cm observations can 
put constraints on reionization models.
\end{abstract}




\section{Introduction}
\label{sec:intro}
The 21-cm signal produced by the hyperfine transition of neutral hydrogen is one of the most promising signals for probing the period from the history of our Universe when it transitioned from a cold and neutral state to a hot and ionised state \citep[e.g.][]{Furlanetto2006CosmologyUniverse,Morales2010ReionizationFluctuations,Zaroubi2013eor}. This transition period is known as the epoch of reionization (EoR), and ended around redshift $z\approx 5.5$--6 \citep{Kulkarni2019Large5.5,Keating2019LongHydrogen,2020MNRAS.494.3080N}.
The 21-cm signal produced during EoR will be redshifted and found at the low frequency end of the radio band. 
It traces the distribution of neutral hydrogen gas present in the intergalactic medium (IGM). Observing this signal will therefore shed light not only on the progress of reionization but also on a range of other issues such as the formation of first compact structures \citep[e.g.][]{Barkana2016TheCosmology}.

The global signal experiment, Experiment to Detect the
Global EoR Signature \citep[EDGES; ][]{bowman2010lower}, has claimed a detection of the sky averaged 21-cm signal from $z\approx 17$ \citep{Bowman2018AnSpectrum}. This detection is debated as its nature questions our current understanding of the early Universe \citep[e.g.][]{Hills2018ConcernsData,Singh2019EDGES}. See \citet{2014PhRvD..90h3522T,Barkana2018PossibleStars,2018PhRvL.121a1101F,2018arXiv180210094M,2018ApJ...858L..17F} and \citet{2019MNRAS.486.1763F} for few possible theories that can explain the \citet{Bowman2018AnSpectrum} observation. Other global experiments, such as the Large-aperture Experiment to Detect the Dark Age \citep[LEDA;][]{price2018design} the Dark Ages Radio Explorer \citep[DARE;][]{burns2017space} and SARAS2 \citep{singh2017} are also trying to detect the global 21-cm signal. As we wait for an independent experiment to confirm the \citet{Bowman2018AnSpectrum} observation, we will assume the standard cosmology in this study.

There also exists numerous efforts to detect the fluctuations in the signal using interferometric radio telescopes, such as the Low Frequency Array \citep[LOFAR; e.g.][]{vanHaarlem2013LOFAR:ARray}, the Murchison Widefield Array \citep[MWA; e.g.][]{Wayth2018mwa} and Hydrogen
Epoch of Reionization Array \citep[HERA; e.g.][]{deboer2017hydrogen}. Even though the fluctuations have not been detected, upper limits have been placed on the 21-cm power spectrum \citep[e.g.][]{Mertens2020ImprovedLOFAR,Trott2020DeepObservations} that have started to put constraints on the reionization process \RefReport{\citep[e.g.][]{Ghara2020ConstrainingObservations,2021arXiv210307483G,Mondal2020TightLOFAR,Greig2020InterpretingObservations,Greig2020ExploringSignal}}. In the near future, the low frequency element of the Square Kilometre Array \citep[SKA-Low; e.g.][]{Mellema2013ReionizationArray,Koopmans2015TheArray} will start observing the 21-cm signal produced during the EoR. SKA-Low is expected to be sensitive enough to not only detect the 21-cm power spectrum but also to produce images \citep[e.g.][]{Mellema2015HISKA,giri2019tomographic}.

The 21-cm signal produced during EoR is highly non-Gaussian and therefore cannot be completely characterised by the 21-cm power spectrum \citep[e.g.][]{Majumdar2018QuantifyingBispectrum}. Therefore we need to explore other summary statistics that are sensitive to the non-Gaussian information contained in the signal. Examples of such summary statistics that have been studied previously are the bispectrum \citep{Majumdar2018QuantifyingBispectrum,Watkinson2019TheHeating}, the position-dependent power spectrum \citep{Giri2019Position-dependentReionization}, size distributions \RefReport{\citep{Kakiichi2017RecoveringTomography,Giri2018BubbleTomography,Giri2019NeutralTomography,2020MNRAS.498.4533B}} and fractal dimensions \citep[][]{Bandyopadhyay2017StudyingDimensions}.

Yet another probe of the non-Gaussian information in the 21-cm signal is a summary statistics that describes the topology of the signal. \RefReport{This topology is quite sensitive to the properties of ioninzing sources \citep[e.g.][]{2021MNRAS.tmp..610H}.}
Already \citet{gnedin2000stagesEoR} used the topology of the \HII (or ionised) regions to heuristically describe the EoR to consist of three stages, namely the \textit{preoverlap}, \textit{overlap} and \textit{post-overlap} stage. Subsequent studies quantified the topology using the genus $g$ or its equivalent the Euler characteristic $\chi$ ($\chi=2-2g$) of the neutral hydrogen distribution \citep[e.g.][]{Gleser2006TheFunctionals,Lee2008TheReionization} or the ionization fraction field \citep{Friedrich2011TopologyReionization}. These works could build on the extensive literature about the use of $\chi$ (or $g$) on the large scale structure \citep[e.g.][]{Gott1986TheUniverse,Gott1992topology,Matsubara1994Genus,MatsubaraTakahikoYokoyama1996GenusFields,Schmalzing1997BeyondStructure,Park2005topology}. 

The Euler characteristic is also equivalent to the third of the Minkowski functionals $V_3$ and some works have considered the full set of these topological quantities \citep[$V_0$--$V_3$,][]{Friedrich2011TopologyReionization, Yoshiura2017StudyingReionization, chen2019stages}. Based on the evolution of the Minkowski functionals the latter study proposed a five stage description of the reionization process.  

The Euler characteristic ($\chi$) for a three-dimensional data set can be written as 
\begin{equation}
    \chi=N_\mathrm{o}-N_\mathrm{t}+N_\mathrm{c}
    \label{eq:basic}
\end{equation}
where the terms on the right hand side are the number of isolated objects, tunnels and cavities. These terms are also known as the Betti numbers. In this paper we explore the information contained in the Betti numbers of the distribution of ionized regions during reionization. It is obvious that they contain more information compared to $\chi$ \citep[e.g.][]{vdW2011AlphaWeb,Chingangbam2012HotandColdSpotCounts,Park2013BettiFields}. Specifically, 
they allow us to study the hierarchical aspect of the topological features of ionised bubbles as reionization proceeds.
We consider how the Betti numbers evolve during the different stages of reionization, how they can describe the state of the IGM and whether the additional information they contain makes them more useful than the more commonly studied Euler characteristic/genus. 

During reionization the distribution of ionised bubbles will imprint its topology on the observable 21-cm signal. Therefore we will also discuss the prospects of measuring the Betti numbers of 3D structures in the upcoming 21-cm observations with SKA-Low, considering both resolution and noise effects. 
\RefReport{As far as we know there exists very few work that considered the Betti numbers in the context of reionization. \citet{Elbers2019PersistentPhenomenology} studied the Betti numbers in heuristic models of reionization which they called the }\HII \RefReport{bubble network. \citet{2018JCAP...10..011K,Kapahtia2019Morphology,2021arXiv210103962K} only studied the Betti numbers} for two dimensional (2D) 21-cm images. However, reionization is a three dimensional (3D) process and SKA-Low will produce 3D data cubes consisting of images at a range of frequencies. We will therefore study the three Betti numbers associated with the 3D topology.





There exists even more metrics to study the topology, such as the persistence homology \citep[e.g.][]{edelsbrunner2008persistent,Zomorodian2005ComputingHomology} that is gaining attention in cosmology \RefReport{\citep[e.g.][]{vdW2010AlphaWeb,vdW2011AlphaWeb,sousbie2011persistentI,sousbie2011persistentII,Pranav2017TheNumbers,Makarenko2018TopologicalDiagrams,2020arXiv201112851W}}. In persistence homology, the evolution of each $\kappa$-dimensional holes in the analysed field is tracked.
\citet{Elbers2019PersistentPhenomenology} have used persistence homology to study the \HII bubble network. We defer the exploration of persistence homology of 21-cm data sets to future studies.


The paper is structured as follows. In the next section, we describe the topological concepts used throughout the paper. Section~\ref{sec:sim_21cm} explains explains our large-scale reionization simulations and the methods used to construct the 21-cm signal from simulation outputs. In Section~\ref{sec:betti_xHII} and \ref{sec:betti_dT}, we discuss the properties and evolution of the Betti numbers of the ionisation and 21-cm signal fields respectively. We discuss the detectability of the Betti numbers in 21-cm observations of upcoming radio telescopes in Section~\ref{sec:21cmImages}. In the final section, we summarise our findings.

\section{Topological measures}
\label{sec:topo_measure}

In this section, we describe the methods we use to measure the topology of our data. In Section~\ref{sec:betti_basics}, we explain the metric we will use to quantify the topology of any field. The fields we analyse in this study are given in form of digital data. Therefore we define an estimator for measuring the topology of digital data in Section~\ref{sec:topo_estimator}. 

\subsection{Betti numbers}
\label{sec:betti_basics}

\begin{figure}
  \centering
\includegraphics[width=0.45\textwidth]{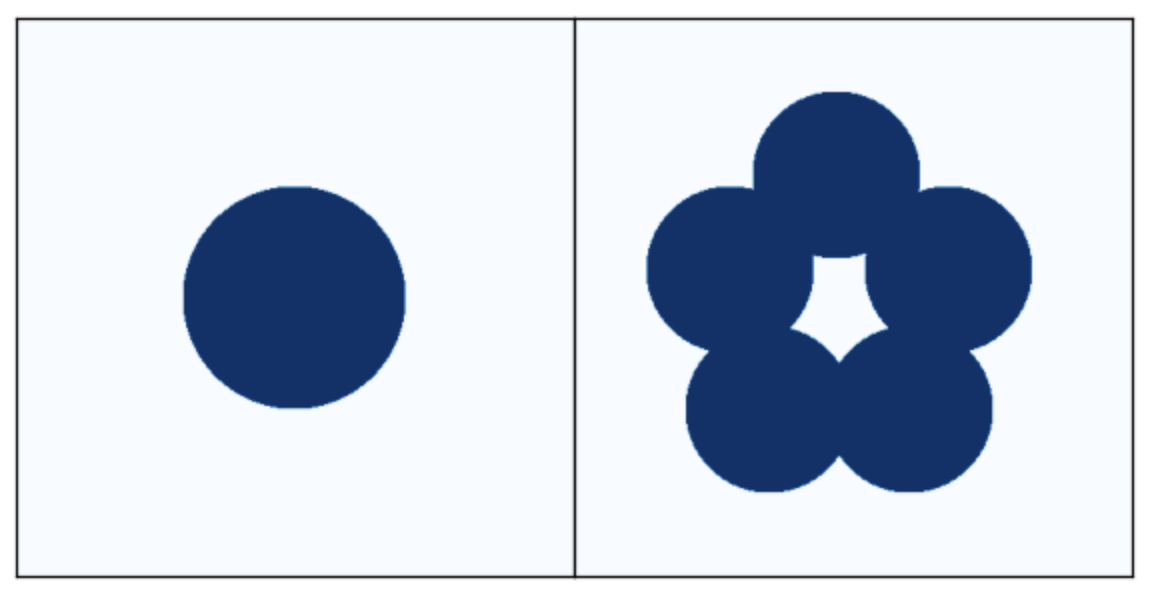}
  \caption{Projection of two different configurations of spheres where the one of the left is hollow. These configurations have Betti numbers $\beta_\kappa=(1,0,1)$ (left) and $\beta_\kappa=(1,1,0)$ (right), giving Euler characteristics $\chi=2$ and $\chi=0$, respectively.}
  \label{fig:cartoon_topo}
\end{figure}

In this study, we quantify the topology of structures in our field with the Betti numbers ($\beta_\kappa$), which are topological invariant quantities \citep{Betti1870SopraDimensioni}. This means they remain unchanged under transformations, such as translation, rotation and deformation. $\beta_\kappa$ is defined as the number of $\kappa$ dimensional holes in the structure we analyse. 
Our data sets are three-dimensional and therefore their topology can be defined with three Betti numbers where each have a simple and intuitive meaning. The zeroth Betti number ($\beta_0$), first Betti number ($\beta_1$) and second Betti number ($\beta_2$) probe the number of isolated objects, tunnels and cavities, respectively.

In its simplest form, the reionization process can be pictured as a network formed by bubbles of ionised hydrogen gas. In Fig.~\ref{fig:cartoon_topo} we show examples of structures formed by spherical bubbles. The left and right panels show projections of 3D data sets containing one and five spherical bubbles respectively.
As they both consist of a single connected structure, both data sets have $\beta_0=1$. The data set in the left panel has no tunnel, hence its $\beta_1=0$. The five bubbles in the right panel form a torus, which contains a tunnel, giving $\beta_1=1$. The projection does not reveal the inner structure of the spheres. If we assume that the single sphere in the left panel has a hollow centre, then this single cavity will give $\beta_2=1$. If we assume that the spheres in the right hand panel are all filled, this gives $\beta_2=0$. 

In previous works, the Euler characteristics ($\chi$), which describes the integral of the Gaussian curvature over the surface of the structure, has been used to study the topology of reionization \citep[e.g.][]{Friedrich2011TopologyReionization,Yoshiura2017StudyingReionization}.
$\chi$ is related to the Betti numbers as  \citep[e.g.][]{Pranav2019TopologyFunctionals},
\begin{eqnarray}
\chi = \beta_0 - \beta_1 + \beta_2 \ .
\label{eq:chi_as_betti}
\end{eqnarray}
(also see Eq.~\ref{eq:basic}). As the Betti numbers are components of $\chi$, two data sets can have the same $\chi$ but different sets of Betti numbers. For example, the data set shown in the right panel of Fig.~\ref{fig:cartoon_topo} has $\beta_\kappa=\{1,1,0\}$ so $\chi=0$, which is the same as for the case when there is no structure. Such a situation is quite plausible in the bubble network produced by reionization and therefore Betti numbers should be better than $\chi$ for distinguishing the topologies of different reionization models.
We refer the readers to \citet{hatcher2002algebraic} and \citet{edelsbrunner2008persistent,edelsbrunner2010computational} for more mathematically rigorous descriptions of these topological quantities.



\subsection{Estimators for digital data}
\label{sec:topo_estimator}

All the fields that we work with in this study are digital data, which is a discrete, discontinuous representation of a given field. In order to capture the topology of the structures in our field, we need to decompose the structures into well defined geometrical components in such a way that the topological properties of the structure are retained. We achieve this by constructing a cubical complex 
\citep[e.g.][]{kaczynski2004computational,Wagner2012}, which is composed of points, line segments, squares and cubes, for any given field. 
We refer the readers to \citet{Wagner2012} and \citet{Giri2019NeutralTomography} for more information about our method to construct and store our cubical complex.

\subsubsection{Betti numbers}
\label{sec:betti_estimator}

We follow the method described in \citet{Gonzalez-Lorenzo2016FastComplexes} to estimate the Betti numbers.
Each isolated object is a group of connected pixels in our data set. We use the connected-component labelling module in \textsc{\small scikit-image} package\footnote{\url{https://scikit-image.org/}} \citep[][]{scikit-image}. We refer inquisitive readers to \citet{fiorio1996two} and \citet{wu2005optimizing} for the algorithm used to label all the connected regions. This method needs less computation time compared to the classical Hoshen-Kopelman algorithm \citep{Hoshen1976PercolationAlgorithm}. 
It is similar to the friends-of-friend algorithm used to find clusters in N-body simulations \citep[e.g.][]{Press1982HowCatalog,Davis1985TheMatter}. The number of regions labelled gives the value of $\beta_0$.
 
From the Alexander duality of any topological complex of three dimensions, we can find the $\beta_2$ by estimating the $\beta_0$ of the complementary field of the data set \citep[e.g.][]{hatcher2002algebraic,Gonzalez-Lorenzo2016FastComplexes}. The remaining first Betti number $\beta_1$ is not calculated directly, but derived from the Euler characteristics ($\chi$) of the data set (as described in Section~\ref{sec:chi_estimator}) according to $\beta_1=\beta_0+\beta_2-\chi$. This method of calculating the $\beta_1$ was introduced by \citet{Gonzalez-Lorenzo2016FastComplexes}.
We have tested our estimator on a Gaussian random field (GRF). The results are presented in Appendix~\ref{sec:topo_grf}.

\subsubsection{Euler characteristics}
\label{sec:chi_estimator}

Once we have constructed the cubical complex for any given digital data, we can estimate the Euler characteristic $\chi$ as,
\begin{eqnarray}
\chi = \sum^{n_\mathrm{d}}_{i=0} (-1)^i \kappa_i \,
\end{eqnarray}
where $n_\mathrm{d}$ is the dimension of the data set. $\kappa_i$ is the number of cubes of dimension $i$ in the constructed cubical complex. In our data set, dimensions of points, line segments, squares and cubes are 0, 1, 2 and 3 respectively. 
See appendix A of \citet{Giri2019NeutralTomography} for more details.

\section{Simulation of 21 cm signal}
\label{sec:sim_21cm}
To test the usefulness of the Betti numbers we make use of simulated data. This section describes our methodology for simulating reionization and the 21-cm signal. Section~\ref{sec:sims} describes the large scale reionization simulations we use and Section~\ref{sec:observed_signal} outlines how we calculate the observable 21-cm signal from them. Finally we explain how we include telescope effects, such as the limited angular resolution and telescope noise in Section~\ref{sec:mock_obs}.

\subsection{Reionization simulations}
\label{sec:sims}

We use two different types of reionization simulations for our study, fully numerical radiative transfer simulations and so-called semi-numerical simulations which rely on sophisticated form of photon counting.

\subsubsection{Fully numerical simulations}
\label{sec:fully_num_sim}

The fully numerical simulations use two steps. We first simulate the evolution of the matter distribution using the $N$-body code \textsc{\small CUBEP$^3$M} \citep{2013MNRAS.436..540H}. The $N$-body simulation is run in a cosmological volume of 714 comoving Mpc (cMpc) with $6912^3$ particles. We use such a large cosmological volume as they are important to capture the large scale fluctuations \citep{Giri2019Position-dependentReionization,Ghara2020ConstrainingObservations} and the large scale topological structure \citep{Giri2019NeutralTomography} and are also a match to the Field of View of SKA-Low.  See \citet{Iliev2014SimulatingEnough} for a discussion about the importance of the size of cosmological volumes for reionization simulations. Our $N$-body simulation run outputs snapshots at each 11.5 megayears (Myrs). We obtain the density field by assigning mass to a cubic grid with 300 cells along each direction by smoothing with a smooth-particle-hydrodynamics (SPH) kernel  \citep{Monaghan1992sph,Iliev2014SimulatingEnough}. 
The cosmological parameters used here are $\Omega_\mathrm{m}=0.27$, $\Omega_\mathrm{k}=0$, $\Omega_\mathrm{b}=0.044$, $h=0.7$, $n_\mathrm{s}=0.96$ and $\sigma_8=0.8$. These values are consistent with the results from \textit{Wilkinson Microwave Anisotropy Probe} (WMAP) \citep{2011ApJS..192...18K} and Planck \citep{PlanckCollaboration2016PlanckHistory,PlanckCollaboration2018PlanckParameters} collaborations.

The reionization process is driven by sources of ionizing photons, which form due to the collapse of baryonic matter in dark matter haloes. We determine the position and mass of dark matter haloes using the spherical overdensity halo-finder discussed in \citet{Watson2013TheAges}. We identify haloes with masses $M_\mathrm{halo} \geq 10^9 \mathrm{M}_\odot$. In these haloes, gas can cool through excitation of atomic hydrogen and they are relatively unaffected by radiative feedback \citep{Sullivan2018UsingReionization}. We follow the convention in \citet{Dixon2016TheReionization} and call these haloes high-mass atomic-cooling haloes (HMACHs). We model the sources by assigning the
ionizing photon production rate, $\dot{N}_\gamma$, to be proportional to the total mass in haloes within that cell, $M$. $\dot{N}_\gamma$ is given as, 
\begin{eqnarray}
\dot{N}_\gamma = \zeta\frac{ M \Omega_\mathrm{b}}{m_\mathrm{p}\Omega_\mathrm{m} }\frac{1}{f_\mathrm{coll}}  \frac{\mathrm{d}f_\mathrm{coll}}{\mathrm{d}t} \,
\label{eq:N_gamma}
\end{eqnarray}
where $\zeta$, $m_\mathrm{p}$ and $f_\mathrm{coll}$ are the source efficiency, mass of proton and total collapsed mass fraction in HMACHs, respectively. 

To simulate the state of gas in the intergalactic medium, we next post-process the snapshots from \textsc{\small CUBEP$^3$M} with the fully numerical radiative transfer code \textsc{\small C$^2$-RAY} \citep{2006NewA...11..374M}. C$^2$RAY employs the short characteristics ray tracing method to solve the radiative transfer equation \citep{raga19993d,lim20033d}.  The ray-tracing is performed up to a comoving distance of 71~Mpc from each source. This limit is meant to approximate the effect of the presence of optically thick absorbers in the IGM that are unresolved in our N-body simulation. This value is consistent with the measurement of the mean free path by \citet{2010ApJ...721.1448S}. For more details about the simulations we refer the reader to \citet{Iliev2006SimulatingReionization}, \citet{Mellema2006SimulatingSignals} and \citet{Dixon2016TheReionization}.


\begin{figure}
  \centering
\includegraphics[width=0.45\textwidth]{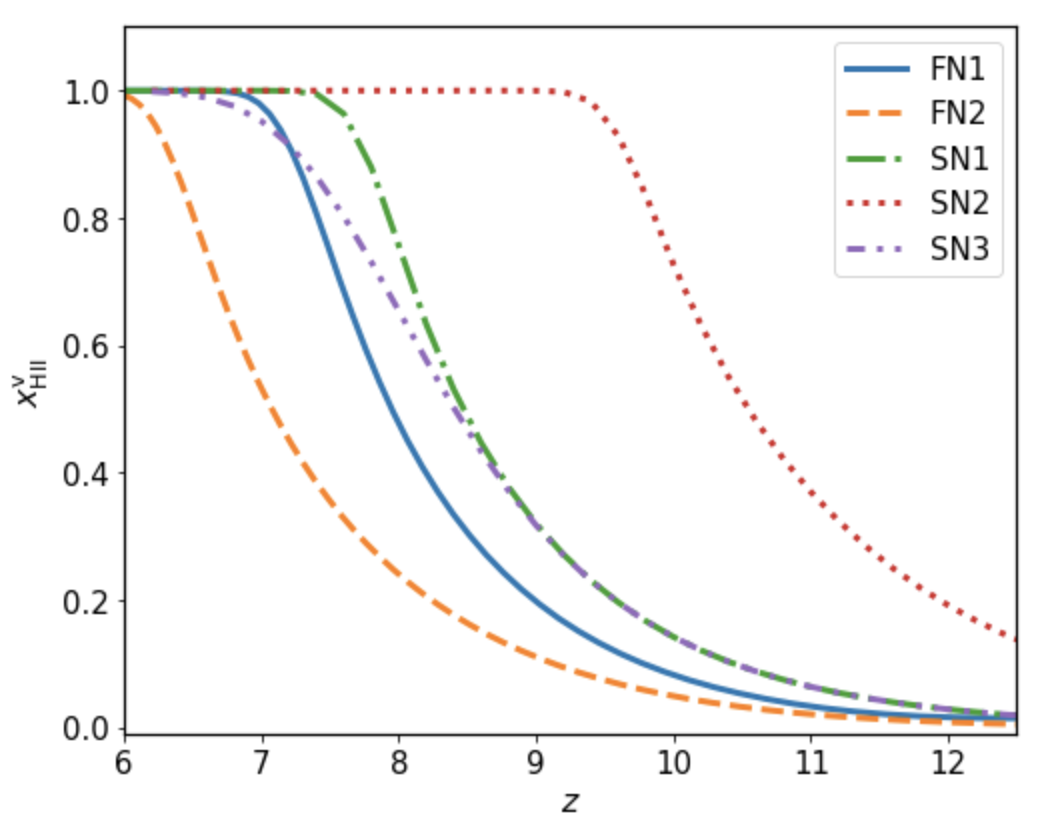}
  \caption{The volume weighted average ionisation fraction of our reionization simulations varying over the redshifts.}
  \label{fig:history_source_models}
\end{figure}

\begin{table*}
	\centering
	\caption{Reionization simulation parameters of various source models. We give the ionizing source parameters, which are ionizing efficiency ($\zeta$), minimum mass ($M_\mathrm{min}$) and mean free path of ionizing photons ($R_\mathrm{mfp}$). The Thompson scattering optical depth ($\tau$) derived from each simulations are also given here.
	}
	\label{tab:sim_list}
	\begin{tabular}{lcccccccc} 
		\hline
		Label  & code & Box size (Mpc) & Mesh & $\zeta$ & $M_\mathrm{min}(M_\odot)$ & $R_\mathrm{mfp}$ (Mpc) & $\tau$ \\
		\hline
		FN1 & \textsc{\small CUBEP$^3$M}+\textsc{\small C$^2$RAY} & 714 & $300^3$ & 50 & $10^9$ & 71 & 0.058 \\
		FN2 & \textsc{\small CUBEP$^3$M}+\textsc{\small C$^2$RAY} & 714 & $300^3$ & 40 & $10^9$ & 71 & 0.050\\
		SN1 & \textsc{\small 21cmFAST} & 714 & $300^3$ & 50 & $10^9$ & 71 & 0.064\\
		SN2 & \textsc{\small 21cmFAST} & 714 & $300^3$ & 50 & $10^8$ & 71 & 0.084\\
		SN3 & \textsc{\small 21cmFAST} & 714 & $300^3$ & 50 & $10^9$ & 20 & 0.062 \\
		\hline
	\end{tabular}
\end{table*}

In this work, we have run two scenarios (FN1 and FN2) using the methods described above. FN1 and FN2 use $\zeta$ values of 50 and 40 respectively. We show the history of these reionization scenarios in Fig.~\ref{fig:history_source_models}. As the sources are less efficient in FN2 compared to the FN1 scenario, reionization is delayed in the latter. The values of the source parameters are listed in Table~\ref{tab:sim_list}.
The Thompson scattering optical depth $\tau$ calculated from these simulations are within the 95 per cent credible interval of the  constraints from \citet{PlanckCollaboration2018PlanckParameters}. The estimated $\tau$ are also given in Table~\ref{tab:sim_list}.


\subsubsection{Semi-numerical simulations}
\label{sec:semi_num_sim}

The fully numerical simulations are computationally expensive. Therefore to study the impact of source parameters on the topology measured with the Betti numbers, we use the fast semi-numerical code \textsc{\small 21cmFAST} \citep[][]{Mesinger201121CMFAST:Signal}. In this code the density field is constructed by evolving the initial density field using second-order Lagrangian perturbation theory \citep[2LPT;][]{scoccimarro1998transients}. \textsc{\small 21cmFAST} builds the ionisation field using the excursion-set approach developed in \citet{2004ApJ...613....1F}. A cell $\pmb{x}$ in the simulation volume is labeled as ionised when it satisfies the following condition,
\begin{eqnarray}
f_\mathrm{coll}(\pmb{x}, R_\mathrm{s}, M_\mathrm{min}) > \zeta^{-1} \ ,
\end{eqnarray}
where $f_\mathrm{coll}(\pmb{x}, R_\mathrm{s}, M_\mathrm{min})$ is the fraction of matter inside a sphere of radius $R_\mathrm{s}$ that has collapsed into haloes of mass larger than $M_\mathrm{min}$ \citep[e.g.][]{Bond1991ExcursionFluctuations,Sheth1999}. $\zeta$ is a again the ionizing efficiency describing the amount of ionizing photons produced by the collapsed mass. This efficiency factor is in principle the same as the one defined in Eq.~\ref{eq:N_gamma}.

We have run three semi-numerical reionization scenarios (SN1, SN2 and SN3) by varying the source and sink parameters in \textsc{\small 21cmFAST}. The values of the parameters are given in Table~\ref{tab:sim_list}. We also show the reionzation history of of these models in Fig.~\ref{fig:history_source_models}. Even though scenario SN1 has the same parameters as FN1, their reionization histories are not identical. This is because SN1 does not explicitly include the photon losses due to recombinations whereas FN1 does. In \textsc{\small 21cmFAST} these losses are assumed to be accounted for in the value of $\zeta$ \citep[see for e.g. eq. 2 in][]{Greig201521CMMC:Signal}.

\subsection{Redshifted 21 cm signal}
\label{sec:observed_signal}

When the 21-cm signal is observed with an interferometry based radio telescope, the recorded observable is known as the differential brightness temperature. The observed differential brightness temperature $\delta T_\mathrm{b}$ of the 21-cm signal, which is given as \citep[e.g.,][]{Mellema2013ReionizationArray},
\begin{eqnarray}
\delta T_\mathrm{b} (\pmb{x},z) \approx 27 x_\mathrm{HI} (\pmb{x},z) \big(1 + \delta (\pmb{x},z)\big)\left( \frac{1+z}{10} \right)^\frac{1}{2}
\left( 1 -\frac{T_\mathrm{CMB}(z)}{T_\mathrm{s}(\pmb{x},z)} \right)\nonumber\\
\left(\frac{\Omega_\mathrm{b}}{0.044}\frac{h}{0.7}\right)
\left(\frac{\Omega_\mathrm{m}}{0.27} \right)^{-\frac{1}{2}} 
\left(\frac{1-Y_\mathrm{p}}{1-0.248}\right)
\mathrm{mK}\ ,
\label{eq:dTb}
\end{eqnarray}
where $x_\mathrm{HI}$ and $\delta$ are the fraction of neutral hydrogen and the density fluctuation respectively. $T_\mathrm{s}$ is the excitation temperature of the two spin states of the neutral hydrogen, known as the spin temperature. $T_\mathrm{CMB}(z)$ is the CMB temperature at redshift $z$ and $Y_\mathrm{p}$ is the primordial helium abundance. 

Here we will work with the assumption of high spin temperature $T_\mathrm{s}\gg$ $T_\mathrm{CMB}$. During the EoR this is a reasonable assumption as even relatively low levels of X-ray radiation can raise the gas temperature above the CMB temperature \citep[e.g.,][]{Pritchard200721-cmReionization,Pritchard201221cmCentury}.
In the high spin temperature limit, $\delta T_\mathrm{b}$ is independent of $T_\mathrm{s}$.

We construct the 21-cm signal at a given redshift $z$ from Eq.~\ref{eq:dTb} using the corresponding three-dimensional neutral fraction and matter density cubes from the \textsc{\small C$^2$-RAY} and \textsc{\small CUBEP$^3$M} simulations.  Each of these 3D data sets represents the characteristics of the signal at the corresponding redshift $z$ and therefore called coeval cubes (CCs). 
We transform these CCs into image cubes by assuming one of the axes to be the frequency axis and the other two position in the sky. This procedure implies that we neglect the two line of sight effects: redshift space distortions \citep[see e.g.][]{Jensen2013ProbingDistortions} and the light cone effect \citep[see e.g.][]{Datta2012Light-coneSpectrum} which would introduce minor distortions along the line of sight. 
We use our publicly available analysis code, {\sc Tools21cm}\footnote{\url{https://tools21cm.readthedocs.io/}} \citep{Giri2020tools21cm}, to simulate the redshifted 21-cm signal cubes.

\subsection{Mock SKA-Low observation}
\label{sec:mock_obs}
In this section, we explain our methods for constructing mock observation with SKA-Low. 

\subsubsection{Limited resolution}
As our simulated 21-cm signal cubes are represented as digital data, they have an intrinsic resolution below which we do not any information. The simulation resolution of our data set is $\Delta x\approx 2.4$ cMpc. This simulation resolution corresponds to different angular scales on the sky at different $z$, which can be given as,
\begin{equation}
\Delta \theta (z) = \frac{\Delta x}{D_\mathrm{c}(z)}\,,
\end{equation}
where $D_\mathrm{c}(z)$ is the comoving distance to redshift $z$. Similarly the frequency scales also vary over redshifts, which can be given as,
\begin{equation}
\Delta \nu = \frac{\nu_0 H(z) \Delta x}{c(1+z)^2}\,,
\end{equation}
where $\nu_0$ is the rest frequency of the 21-cm line, $H(z)$ the Hubble parameter at redshift $z$ and $c$ the speed of light. For $z=8$, these equations give values $\Delta \theta=0.881$ arcmin and $\Delta \nu=0.133$~MHz.

Apart from the simulation resolution, we will also be limited by the resolution of the telescope. As SKA-Low is an interferometry based radio telescope, it observes the sky in so-called $uv$ space, which corresponds to the Fourier transform of the sky. We refer the readers to \citet{rohlfs2013tools} for details about radio telescope observations. 

A radio interometric telescope comprises of radio antennae placed at different locations. Each antennae pair forms a baseline and it observes the signal at an angular scale proportional to the length of the baseline. For observing the 21-cm signal this angular scale can be written as,
\begin{equation}
\Delta \theta_\mathrm{telescope} (z) = \frac{21(1+z)}{B}\,
\label{eq:angluar_res_telescope}
\end{equation}
where $B$ is the length of the baseline in cm. The first generation of SKA-Low will be sparsely filled with antennae outside a diameter of 2 km around the centre \citep[e.g.][]{Mellema2013ReionizationArray}. Therefore a 21-cm data set produced with baselines longer than 2 km will be quite noisy. In this work we will approximate the synthesized beam of SKA-Low by a Gaussian with a full-width half maximum (FWHM) given by Eq.~\ref{eq:angluar_res_telescope} with $B=2$~km. For $z=8$ this gives 
$\Delta \theta_\mathrm{telescope} \approx 3$~arcmin.

An interferometric radio telescope also has a smallest baseline which sets the largest scales that it can image. To implement this effect, we subtract its average value from each slice of the 21-cm signal data set along the frequency (or redshift) axis. This effect means that interferometric images lack absolute calibration of the signal.
Finally, to attain isotropic resolution in our data set we convolve the signal along the frequency axis with a tophat filter of the same physical width as the Gaussian beam. For $z=8$ this gives 
$\Delta \nu \approx 0.5$~MHz.

\subsubsection{System noise}
\label{sec:noise_model}

\begin{table}
	\centering
	\caption{The parameters used in this study to model the telescope properties.}
	\label{tab:telescope_param}
	\begin{tabular}{lccccc} 
		\hline
		Parameters & Values \\
		\hline
        System temperature ($T_\mathrm{sys}$) & $60 (\frac{\nu}{300\mathrm{MHz}})^{-2.55}$ K  \\
		Effective collecting area ($A_\mathrm{D}$) & 962 $\mathrm{m}^2$  \\
        Critical frequency ($\nu_\mathrm{c}$) & 110 MHz \\
        Declination & -30$^\circ$ \\
        Observation hour per day & 6 hours \\
        Signal integration time & 10 seconds \\
		\hline
	\end{tabular}
\end{table}


The 21-cm observations with SKA-Low will also be prone to instrumental noise.
In order to simulate this noise, we follow the approach in \citet{Ghara2017ImagingSKA} and \citet{Giri2018OptimalObservations}.
We model the effects using the currently available configuration\footnote{The configuration is described in document SKA-TEL-SKO-0000557 Rev 1 and the positions of the individual stations in document SKA-TEL-SKO-0000422 Rev 2, both retrievable at \url{https://astronomers.skatelescope.org/documents/}} of the first phase of SKA-Low, 
which has 512 antennae stations each with a diameter of 35 m. 
We present the telescope parameters that we use in this study in Table~\ref{tab:telescope_param}.

We explore two observation time ($t_\mathrm{obs}$). One is an optimistic case of $t_\mathrm{obs}$ = 1000 h and other is a pessimistic case of $t_\mathrm{obs}$ = 100 h.
At simulation resolution, the system noise at 1000 h and 100 h will be very high. As the this noise is uncorrelated while the signal is correlated, the signal-to-noise ratio (SNR) will go up when the resolution is decreased. As discussed earlier, SKA-Low will have a limited resolution defined by the maximum baseline. Previous studies have shown that we can get interesting image data with 1000 h observation at a resolution corresponding to $B=2$ km \citep[e.g.][]{Mellema2015HISKA,Ghara2017ImagingSKA,Giri2018OptimalObservations, Giri2019NeutralTomography}. However the SNR will be much lower for a 100 h observation at the same resolution. To achieve a similar SNR, we reduce the resolution by a factor of 2 for the 100 h mock observations.

We would like to note that the noise model we use assumes perfect calibration. In reality it is not always possible to reach this theoretical noise level. Therefore To achieve the same SNR values as in our example with the real SKA-Low might require longer observation times or use of a lower resolution.


\section{Betti numbers of ionisation field}
\label{sec:betti_xHII}

The formation, merger and evolution of ionised bubbles during reionization results in a complex morphology of \HII regions.
In this section we investigate the Betti numbers estimated from the ionisation field in our simulations.

\subsection{Evolution during reionization}
\label{sec:evol_betti}

\begin{figure}
  \centering
  \includegraphics[width=0.46\textwidth]{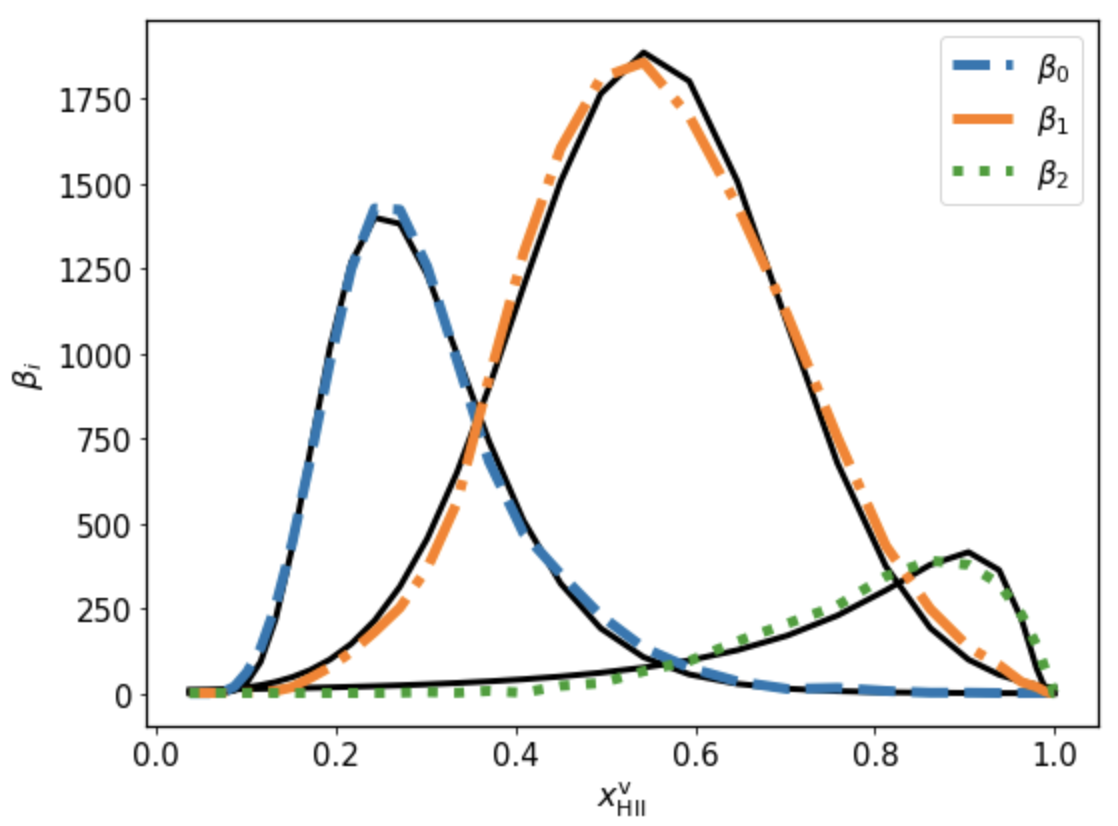}
  \caption{Evolution of the Betti numbers estimated from the $x_\mathrm{HII}$ field of the FN1 simulation. The black solid lines are the fits to the curves. See text for a description of these fits and Table~\ref{tab:fitted_params} for the fit parameters.}
  \label{fig:betti_SimRes}
\end{figure}

\begin{figure}
  \centering
  \includegraphics[width=0.46\textwidth]{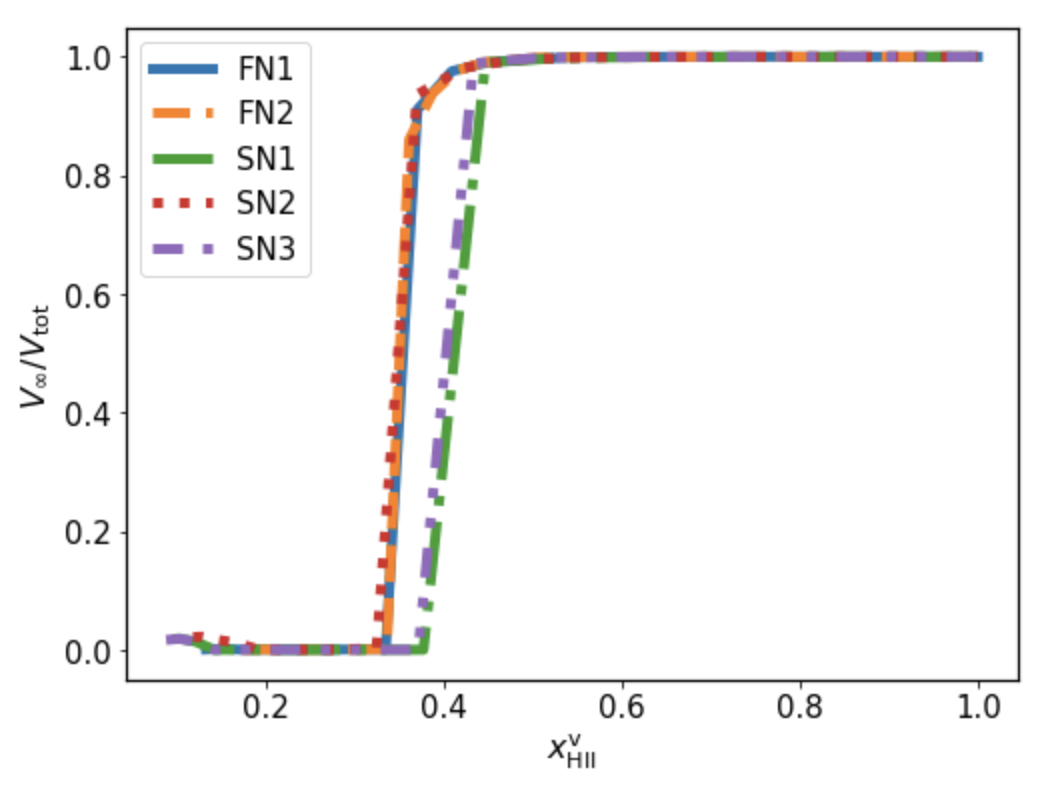}
  \caption{Percolation curves for the ionized percolation cluster. The number on the ordinate, $V_\infty/V_\mathrm{tot}$, is the fraction of the total ionized volume contained in the percolation cluster. The different curves show the evolution of this quantity for our five different reionization scenarios.}
  \label{fig:perc_curve}
\end{figure}

Fig.~\ref{fig:betti_SimRes} shows the evolution of Betti numbers estimated from the ionisation field of our FN1 simulations by defining structures using a threshold of $x_\mathrm{HII}=0.5$. These Betti numbers trace the evolution topology of \HII regions during reionization. \RefReport{Here and in the remainder of this paper we plot the evolution of the Betti numbers against the fraction of our simulation volume that has been ionized, rather than time or redshift. Using the ionized volume fraction $x_\mathrm{HII}^\mathrm{V}$ connects the evolution of the Betti numbers to the different stages of reionization and also makes it easier to compare the results of different reionization scenarios in which reionization may have evolved differently.}

When reionization starts we see an increase in $\beta_0$. This increase in $\beta_0$ is due to the appearance of ionised bubbles around the sources emitting ionising photons. As time passes, more sources form and ionise the hydrogen gas in the IGM around them.  
In the initial stages of reionization we not only see appearance of new ionised bubbles but also overlap of bubbles. A comparison between the value of $\beta_0$ and the number of sources shows that such overlap occurs right from the start of reionization. The total number of bubbles will decrease when multiple ionised bubbles overlap. When the rate by which \HII regions start to overlap becomes larger than the formation rate of new isolated ionised bubbles, the value of $\beta_0$ will start to decrease. In our model, we observe this turn-around at $x^\mathrm{v}_\mathrm{HII}\approx 0.2$. The value of $\beta_0$ keeps decreasing as reionization proceeds and converges to unity at around $x^\mathrm{v}_\mathrm{HII}\approx 0.7$. 

At $x^\mathrm{v}_\mathrm{HII}\approx 0.1$, $\beta_1$ starts to increase. $\beta_1$ measures the number of tunnels and these tunnels can be formed by overlapping bubbles (see Section~\ref{sec:betti_basics} for a simple example). Therefore the growth of $\beta_1$ is a sign of substantial bubble overlap modulating the topology of the ionisation field. As reionization progresses, the number of tunnels increase until it reaches a maximum value beyond which $\beta_1$ starts to decrease. In the FN1 simulation this peak is reached around $x^\mathrm{v}_\mathrm{HII}\approx 0.55$. Unlike $\beta_0$, the value of $\beta_1$ remains much larger than 1 until the end of reionization.

Cavities inside \HII regions start forming during the middle stages of reionization, as indicated by the slow rise of $\beta_2$, starting around $x^\mathrm{v}_\mathrm{HII}\approx 0.2$. We can visualise these cavities as isolated neutral regions. As reionization progresses initially large connected neutral regions break up and form multiple isolated neutral regions. The value of $\beta_2$ attains a maximum around $x^\mathrm{v}_\mathrm{HII}\approx 0.9$ after which it falls to zero once reionization completes. In any inside-out reionization scenario such as FN1, these later isolated neutral regions trace the cosmic voids. \citet{Giri2019NeutralTomography} noted that the number of isolated neutral regions in the later stages of reionization are relatively rare compared to the number of isolated ionised regions at the start of reionization. This is confirmed by comparing the maximum values of the $\beta_0$ and $\beta_2$ curves, which shows that the latter is about 3.5 times lower.

From the three curves we obtain three clear maxima, as well as the crossing points between $\beta_0$ and $\beta_1$ as well as between $\beta_1$ and $\beta_2$. As $\beta_1$ enters the calculation of the Euler characteristic $\chi$ with a different sign than $\beta_0$ and $\beta_2$, the above evolutionary patterns cannot be distinguished when considering $\chi$. It is for example impossible to separate the early rise of the number of tunnels and the decrease of the number of isolated regions due to overlap.
At $x^\mathrm{v}_\mathrm{HII}\approx 0.35$ $\beta_0\approx\beta_1$. However the small positive contribution from $\beta_2$ at that time will push the location of $\chi=0$ to a slightly later stage.

\subsection{Connection between Betti numbers and percolation}
These crossing points of Betti curves appear to be connected to the formation of the percolation clusters. \citet{Furlanetto2016ReionizationTheory} showed that a large network of overlapping ionised bubbles spanning the entire simulation volume forms during the early stages of reionization. This network is known as the percolation cluster. A similar percolation event\footnote{This should perhaps be called a depercolation event as the neutral percolation cluster disappears.} takes place towards the end of reionization when the last network of neutral regions spanning the entire volume disappears. As pointed out by \citet{Furlanetto2016ReionizationTheory} this type of percolation behaviour is a hallmark of phase transitions.

Fig.~\ref{fig:perc_curve} shows the percolation curves, i.e.\ the volume fraction of \HII regions contained in the percolation cluster, for all our simulations. Here we focus on the percolation curve of the FN1 scenario, the percolation curves of other scenarios are discussed in Section~\ref{sec:source_models_xHII}. We see that the appearance of the percolation cluster coincides with the time when \bettiOne surpasses \bettiZero. Likewise we find that the disappearance of the neutral percolation cluster coincides with the time when \bettiOne$\approx~$\bettiTwo. It was recently proposed by \RefReport{\citet{PhysRevE.100.032414} and} \citet{Bobrowski2020HomologicalCharacteristic} that such a connection between Betti numbers and percolation may be universal.

\subsection{Parametrization of the evolution of Betti numbers}
The evolution curves for each of the Betti numbers for scenario FN1 have strikingly regular shapes. After inspection of the other scenarios we find that the shapes appear to be independent of the source model driving the reionization process. We will discuss these additional reionization scenarios in Section~\ref{sec:source_models_xHII}. The curves for \bettiZero and \bettiTwo seem to match a log-normal shape, whereas \bettiOne appears to have a Gaussian shape. We therefore propose the following parametric forms for each Betti number,
\begin{eqnarray}
\beta_0(x^\mathrm{v}_\mathrm{HII}) = A_0 \exp{\left[-\frac{1}{2}\left(\frac{\ln x^\mathrm{v}_\mathrm{HII} - \ln \mu_0}{\sigma_0} \right)^2 \right]} \ ,
\label{eq:fit_b0}
\end{eqnarray}
\begin{eqnarray}
\beta_1(x^\mathrm{v}_\mathrm{HII}) = A_1 \exp{\left[-\frac{1}{2}\left(\frac{x^\mathrm{v}_\mathrm{HII} - \mu_1}{\sigma_1} \right)^2 \right]} \ ,
\label{eq:fit_b1}
\end{eqnarray}
\begin{eqnarray}
\beta_2(x^\mathrm{v}_\mathrm{HII}) = A_2 \exp{\left[-\frac{1}{2}\left(\frac{\ln (1-x^\mathrm{v}_\mathrm{HII}) - \ln \mu_2}{\sigma_2} \right)^2 \right]} \ ,
\label{eq:fit_b2}
\end{eqnarray}
where $A_i$, $\mu_i$ and $\sigma_i$ are the fitting parameters. $A_i$ models the height of the curves. $\mu_i$ and $\sigma_i$ define the peak position and width of the curves. We show the fits to the Betti numbers curve with black solid lines in Fig.~\ref{fig:betti_SimRes}. The fit parameters are listed in Table~\ref{tab:fitted_params}. \RefReport{Here and in the rest of this paper we use 10-fold cross validation to estimate the goodness of the fit \citep[e.g.][]{hastie2009elements,2020MNRAS.491.5277G}.}

The fact that simple functional forms can be used to describe the evolution of the Betti numbers for reionization is another advantage over the Euler characteristic for which no simple fit appears to exist. 


\subsection{Dependence on simulation resolution}
\label{sec:dependence_SimRes}

\begin{figure*}
  \centering
  \includegraphics[width=0.98\textwidth]{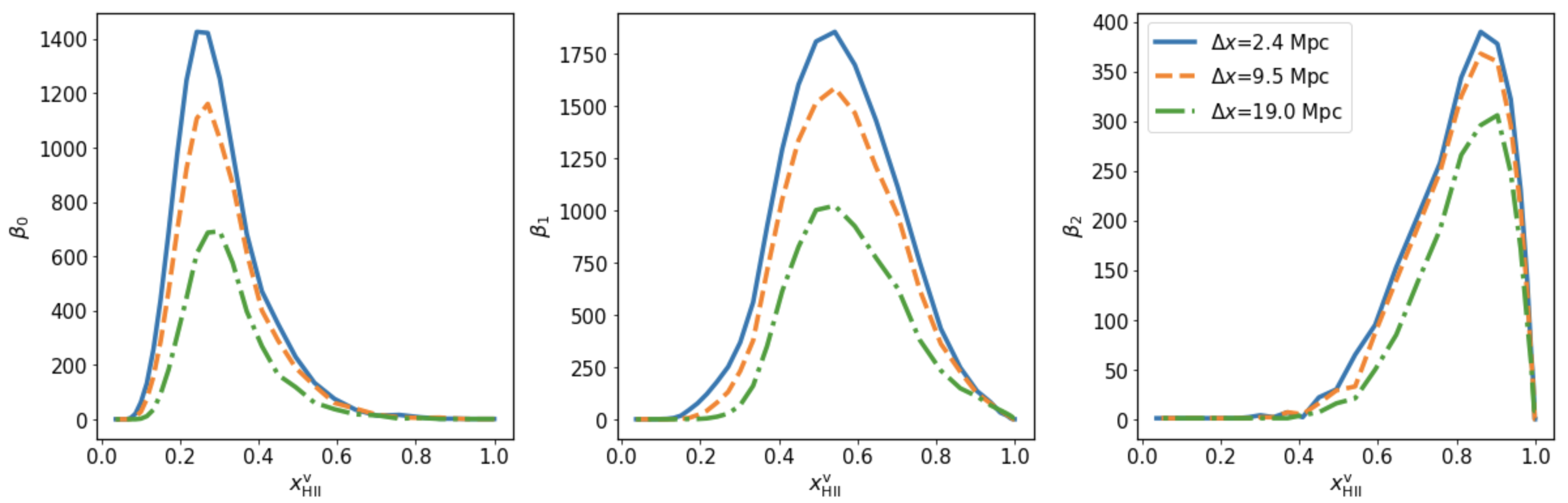}
  \caption{Evolution of the Betti numbers estimated from the $x_\mathrm{HII}$ field of the FN1 simulation at various resolutions. The solid, dashed and dash-dotted line-styles give Betti numbers estimated from data set at simulation resolution, 4-cell smoothing and 8 cell smoothing respectively. The latter two approximately correspond to the resolution achieved by SKA-Low observations with a maximum baseline of 2 km and 1 km respectively for redshift $z\approx 9$.}
  \label{fig:betti_xHII_vary_res}
\end{figure*}

The topology of \HII regions is sensitive to the resolution of our simulation volume. \citet{Friedrich2011TopologyReionization} showed the impact of simulation resolution on the Euler characteristics. In this section, we study how the Betti numbers depend on the simulation resolution.

We show the Betti numbers curves estimated from the FN1 simulation volumes at different resolutions in Fig.~\ref{fig:betti_xHII_vary_res}. The solid curve represents the Betti numbers estimated from the data set at the intrinsic resolution of our simulation ($\Delta x \approx 2.4$ Mpc). The dashed and dash-dotted lines represent their values estimated after 4-cell and 8-cell smoothing respectively. The 4-cell and 8-cell smoothing approximately correspond to the resolution of SKA-Low observations at $z\approx 9$ when maximum baselines of 2 km and 1 km are used respectively.

When we fit these curves with the parametric form given in Eq.~\ref{eq:fit_b0}, \ref{eq:fit_b1} and \ref{eq:fit_b2}, we find that the peak positions and widths of the curve do not change as the resolution degrades. The resolution only affects the amplitude parameter $A_i$ for all the Betti numbers curve. This decrease in amplitude is expected as the number of features contained in the data set decreases when the resolution is degraded. The fact that the peak positions and widths are less affected adds to the attractiveness of the Betti numbers as topological quantities.

\subsection{Comparing the topology of different reionization scenarios}
\label{sec:source_models_xHII}

\begin{figure*}
  \centering
  \includegraphics[width=0.98\textwidth]{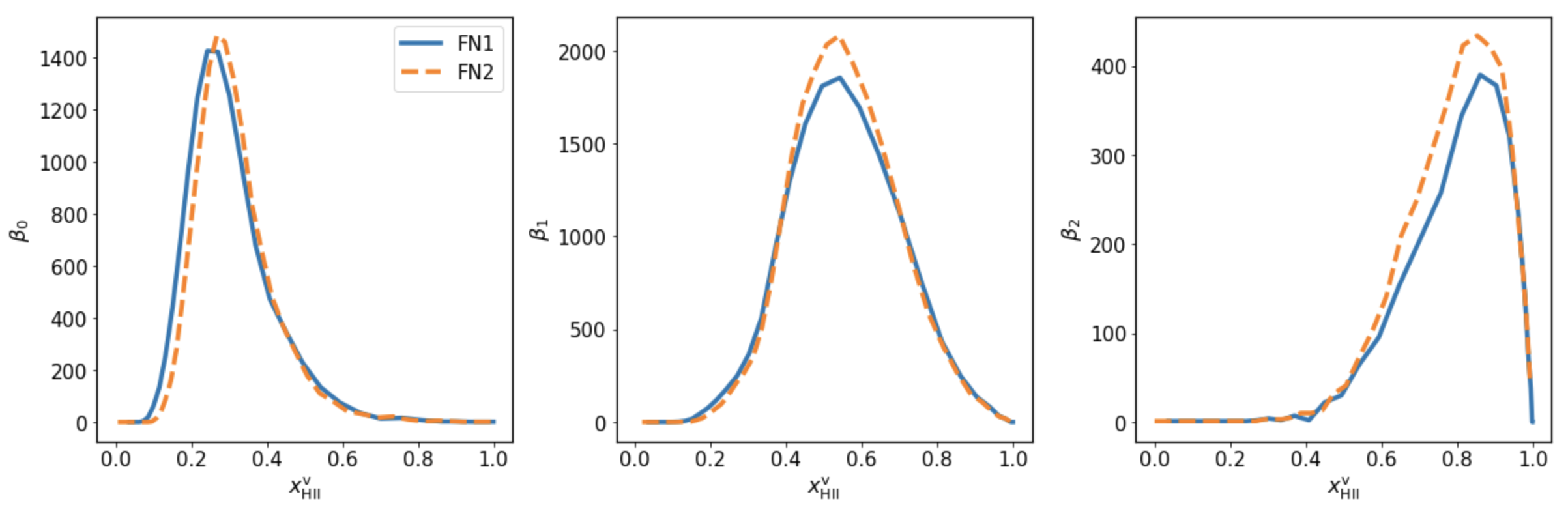} \\
  \includegraphics[width=0.98\textwidth]{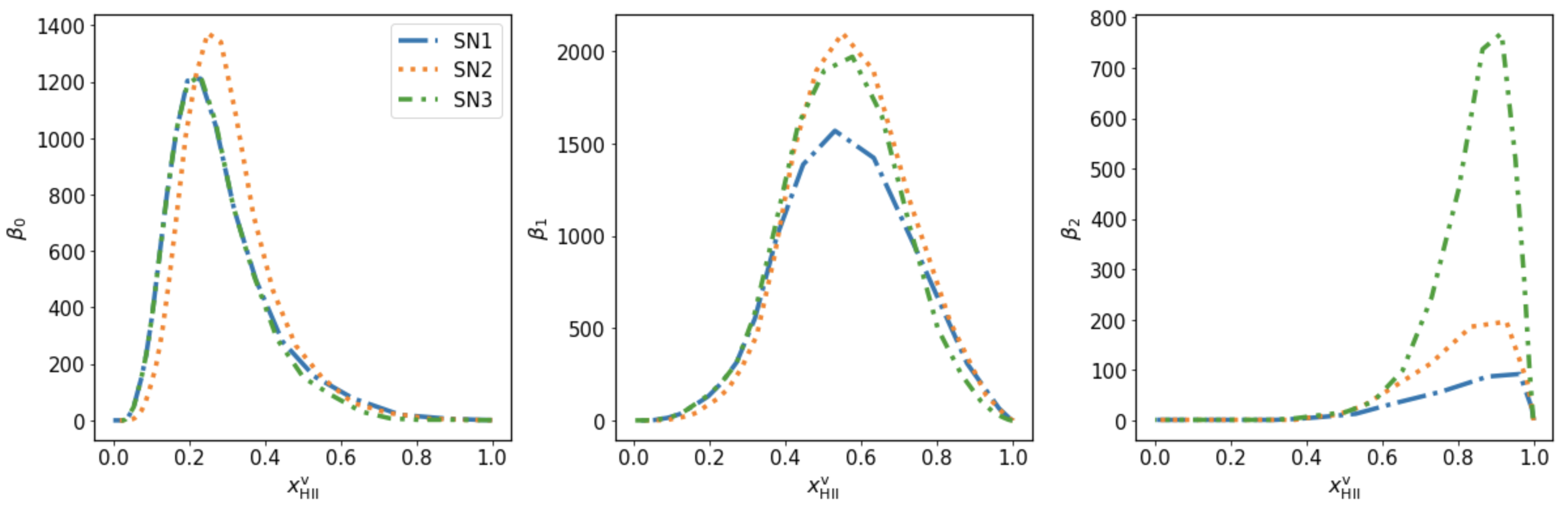}
  \caption{Evolution of the Betti numbers estimated from the $x_\mathrm{HII}$ field of different simulations. The top and bottom panels show the Betti numbers for our fully numerical and semi-numerical simulation models respectively.}
  \label{fig:betti_xHII_models}
\end{figure*}

\begin{table*}
\caption{The values of parameters when we model the Betti numbers using Eq.~\ref{eq:fit_b0}, \ref{eq:fit_b1} and \ref{eq:fit_b2}. The fit parameters for $x_\mathrm{HII}$ are Betti number curves estimated from the field at simulation resolution. The fit parameters for $\delta T_\mathrm{b}$ are Betti number curves estimated from the mock SKA-Low 21-cm images. We give the fit parameters corresponding to mock observations at $t_\mathrm{obs}=1000$ h and $B=2$ km. The fit parameters given in brackets correspond to mock observations at $t_\mathrm{obs}=100$ h and $B=1$ km.
}
\begin{center}
\begin{tabular}{ccccccccp{1cm}p{1cm}p{1cm}p{1cm}p{1cm}p{1cm}p{1cm}}
    \hline
    {} & {\bf Fit} & {} & $\pmb{x_\mathrm{HII}}$ & {} & {} & $\pmb{\delta T_\mathrm{b}}$ & {}\\
    \textbf{Models} & {\bf parameters} & $\beta_0$ & $\beta_1$ & $\beta_2$ & $\beta_0$ & $\beta_1$ & $\beta_2$\\
    \hline
    \hline
    \\
    \multirow{3}{*}{FN1} & $A_i$ & 1409 & 1886 & 415 & 50 (\CommentOut{32}--) & 512 (443) & 92 (80) \\
    & $\mu_i$ & 0.25 & 0.55 & 0.10 & 0.27 (\CommentOut{0.11}--) & 0.61 (0.61) & 0.04 (0.03) \\
    & $\sigma_i$ & 0.71 & 0.15 & 0.43 & 0.55 (\CommentOut{0.24}--) & 0.16 (0.18) & 0.33 (0.34)
    \vspace{0.3cm}\\
    \multirow{3}{*}{FN2} & $A_i$ & 1443 & 2092 & 476 & 156 (\CommentOut{63}--) & 562 (500) & 90 (80)\\
    & $\mu_i$ & 0.27 & 0.55 & 0.11 & 0.29 (\CommentOut{0.31}--) & 0.63 (0.62) & 0.04 (0.03) \\
    & $\sigma_i$ & 0.75 & 0.14 & 0.46 & 0.72 (\CommentOut{0.66}--) & 0.15 (0.18) & 0.38 (0.36)
    \vspace{0.3cm}\\
    \multirow{3}{*}{SN1} & $A_i$ & 1195 & 1652 & 112 & 161 (\CommentOut{21}--) & 470 (371) & 43 (45) \\
    & $\mu_i$ & 0.21 & 0.56 & 0.06 & 0.24 (\CommentOut{0.17}--) & 0.61 (0.63) & 0.03 (0.05) \\
    & $\sigma_i$ & 0.63 & 0.16 & 0.37 & 0.69 (\CommentOut{0.37}--) & 0.16 (0.15) & 0.26 (0.45)
    \vspace{0.3cm}\\
    \multirow{3}{*}{SN2} & $A_i$ & 1351 & 2168 & 223 & 66 (\CommentOut{14}--) & 461 (342) & 43 (25) \\
    & $\mu_i$ & 0.25 & 0.56 & 0.08 & 0.31 (\CommentOut{0.32}--) & 0.60 (0.64) & 0.02 (0.05) \\
    & $\sigma_i$ & 0.69 & 0.15 & 0.41 & 0.64 (\CommentOut{0.33}--) & 0.19 (0.16) & 0.19 (0.25)
    \vspace{0.3cm}\\
    \multirow{3}{*}{SN3} & $A_i$ & 1206 & 2048 & 800 & 138 (\CommentOut{25}--) & 689 (688) & 130 (200) \\
    & $\mu_i$ & 0.21 & 0.55 & 0.09 & 0.24 (\CommentOut{0.10}--) & 0.61 (0.67) & 0.03 (0.05) \\
    & $\sigma_i$ & 0.64 & 0.15 & 0.48 & 0.68 (\CommentOut{0.26}--) & 0.16 (0.11) & 0.19 (0.45) \\
    \hline
\end{tabular}
\end{center}
\label{tab:fitted_params}
\end{table*}

In this section we compare the topology of various reionization models using the Betti numbers. The top panels of Fig.~\ref{fig:betti_xHII_models} show the Betti curves for our fully numerical simulations. The FN1 and FN2 simulations use source efficiencies $\zeta=50$ and 40 respectively. Therefore reionization is delayed in FN2 (see Fig.~\ref{fig:history_source_models}). However, even though the reionization history is different the peak positions of the Betti curves are quite similar. The $\beta_0$ gives the number of isolated ionised regions and at early times, this number will be proportional to the number of active ionising sources and not the efficiency of those sources. Note however that $\beta_0$ will not be exactly equal to the number of sources as some regions are formed by clustered sources. Fig.~\ref{fig:history_source_models} shows that FN2 reaches a similar stage $\Delta z\approx 1$ later than FN1. The number density of ionising sources will not differ that much between those two epochs. Therefore the $\beta_0$ curves for both these models are quite similar. However, the topology does differ, as evidenced by the \bettiOne and \bettiTwo curves. The FN2 simulation contains more tunnels during the middle stages of reionization and more cavities from about the middle until near the end of reionization. Using all the curves together, we can distinguish between these two models.

To test the robustness of our conclusions, we also study the topology of reionization in a set of semi-numerical simulations. As semi-numerical simulations are fast, we can easily vary the simulation parameters and observe its impact on the Betti numbers. In the bottom panel of Fig.~\ref{fig:betti_xHII_models}, we show the Betti numbers curves for our three semi-numerical simulations (see Table~\ref{tab:sim_list}). 
SN1 and SN3 use the same $M_\mathrm{min}$ value ($10^9 M_\odot$) for dark matter halos, which means that reionization is driven by the same source population. Therefore the $\beta_0$ curves for SN1 and SN3 overlap with each other. The $M_\mathrm{min}$ used in SN2 is $10^8 M_\odot$. As a result SN2 contains more ionising sources compared to SN1 and SN3. We see the imprint of this on the $\beta_0$ curve, which reaches a higher amplitude for the SN2 simulation.  

The SN1 and SN2 simulations assume $R_\mathrm{mfp}$ to be 71 Mpc where as SN3 simulation assumes it to be 20 Mpc. SN1 and SN3 otherwise have the same source parameters. $R_\mathrm{mfp}$ describes the largest distance ionising photons can travel from a source. At the early times, the value of $R_\mathrm{mfp}$ does not have a significant impact on the topology as most \HII regions are smaller than $R_\mathrm{mfp}$. However, once larger regions form during the middle and late stages of reionization, a smaller value for $R_\mathrm{mfp}$ will inhibit the growth of ionized regions. This will allow more tunnels to survive, as evidenced by the difference between the \bettiOne curves of SN1 and SN3. The largest impact of $R_\mathrm{mfp}$ is associated with the final phases of reionization as can be seen from the $\beta_2$ curves. The values are much higher for SN3, indicating that imposing a short mean free path leads to the formation of many more small neutral islands.



As pointed out in Section~\ref{sec:betti_xHII}, the rise of \bettiOne in simulation FN1 marks the appearance of the percolation cluster. The percolation curves for all our reionization models can be found in Fig.~\ref{fig:perc_curve}. 
We find that for all models, percolation happens roughly at the epoch when \bettiOne crosses \bettiZero confirming our hypotesis.


The Betti numbers curves from all our reionization models, both fully numerical and semi-numerical, show the characteristic forms defined in Section~\ref{sec:betti_xHII}. We present the values of the fit parameters for all models in Table~\ref{tab:fitted_params}. We conclude that these fitting formulae
Eq.~\ref{eq:fit_b0}, \ref{eq:fit_b1} and \ref{eq:fit_b2} provide a useful description of the evolution of Betti numbers in reionization simulations.

\section{Betti numbers of 21-cm signal}
\label{sec:betti_dT}

\begin{figure*}
  \centering
  \includegraphics[width=1.\textwidth]{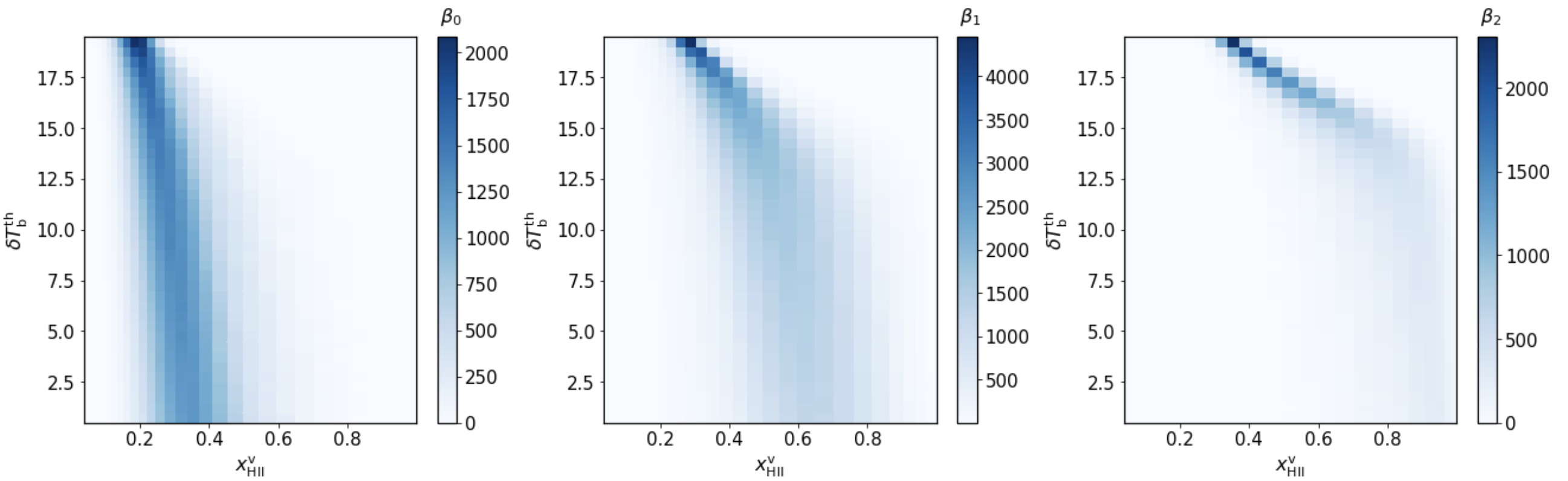}
  \caption{Contour plots of the values of $\beta_0$ (left panel), $\beta_1$ (middle panel) and $\beta_2$ (right panel) of the $\delta T_\mathrm{b}$ field from the FN1 simulation as a function of global ionization fraction  $x^\mathrm{v}_\mathrm{HII}$ and threshold value ($\delta T^\mathrm{th}_\mathrm{b}$).}
  \label{fig:betti_dT_vary_xs_thres}
\end{figure*}

In this section we will study the evolution of Betti numbers estimated from the 21-cm signal ($\delta T_\mathrm{b}$). As explained in Section~\ref{sec:observed_signal}, this signal depends on neutral fraction $\big( x_\mathrm{HI}(\pmb{r}) = 1-x_\mathrm{HII}(\pmb{r}) \big)$, density and spin temperature fields (see Eq.~\ref{eq:dTb}). For our study we assume the Universe has been heated before reionization begins causing the spin temperature factor to saturate to 1 as $\big( T_\mathrm{s}\gg T_\mathrm{CMB} \big)$. Even with this assumption, the $\delta T_\mathrm{b}$ will still contain properties of both the  ionisation and density fields.

We first study the topology of structures by putting a threshold ($\delta T^\mathrm{th}_\mathrm{b}$) on the $\delta T_\mathrm{b}$ field. We measure the topology of structures with pixel values below a given value of $\delta T^\mathrm{th}_\mathrm{b}$.
In Fig.~\ref{fig:betti_dT_vary_xs_thres}, we show colour plots of the Betti numbers estimated from the $\delta T_\mathrm{b}$ field of the FN1 simulation against the values of $\delta T^\mathrm{th}_\mathrm{b}$ (y-axis) and $x^\mathrm{v}_\mathrm{HII}$ (x-axis). 
For high values of $\delta T^\mathrm{th}_\mathrm{b}$, the topology is mostly defined by the fluctuations in the density field. However the Betti numbers will not exactly trace the topology of the density field as reionization will reduce the signal in the (predominantly) dense regions. 
For low $\delta T^\mathrm{th}_\mathrm{b}$ value, the structures trace the topology of ionisation field. For $\delta T^\mathrm{th}_\mathrm{b} \lesssim 10$ mK, the Betti numbers estimated from $\delta T_\mathrm{b}$ fields are similar to that estimated from $x_\mathrm{HII}$ field of FN1 simulation.

For different reionization scenarios we will obtain different colour plots from the one shown in Fig.~\ref{fig:betti_dT_vary_xs_thres}. However, it is difficult to compare scenarios with these plots as they will be affected by instrumental limitations such as lack of an absolute calibration of the signal, limited resolution and noise. In the next section, we propose a method to explore the topological properties of reionization using mock 21-cm observations for the future SKA-Low telescope.

\section{Betti numbers of 21-cm images observed with radio telescope}
\label{sec:21cmImages}

As discussed in the previous section, the structures in the ionisation field will be imprinted on the 21-cm signal. Therefore we can study the morphology of \HII regions if we can identify these regions in 21-cm images. As the future SKA-Low has been designed to be able to produce such images \citep{Mellema2015HISKA}, we will consider mock SKA-Low observations in this section and study our ability to estimate the Betti numbers from such observations. 

Identifying the ionised (or neutral) regions in 21-cm images is a non-trivial problem. \citet{Giri2018OptimalObservations} compared various methods taken from the field of image processing and found that the superpixel method works the best in identifying ionised (or neutral) regions in 21-cm images. In the superpixel method, connected pixels with similar properties are grouped together. These pixel groups are known as superpixels. We then construct the histogram from the mean intensity values of all the superpixels. From this histogram we find the threshold for the superpixels that classifies them into ionised and neutral superpixels. 
See \citet{Giri2018OptimalObservations} for a detailed description of the method. \RefReport{Note that the identification of the ionized regions at the same time yields an estimate for the fraction of the observed volume which is ionized, thus allowing to determine the Betti numbers as a function of ionized volume fraction rather than redshift.}

In Section~\ref{sec:lim_res}, we consider mock SKA-Low observations with infinite observation time as we want to first learn what we can achieve in this most optimal case with our method. In this situation, the 21-cm images will be free from noise and only be affected by the limited resolution and the lack of absolute calibration. In Section~\ref{sec:tele_noise} we study the impact of telescope noise on our analysis strategy.

\subsection{Limited resolution}
\label{sec:lim_res}

\begin{figure*}
  \centering
  \includegraphics[width=0.98\textwidth]{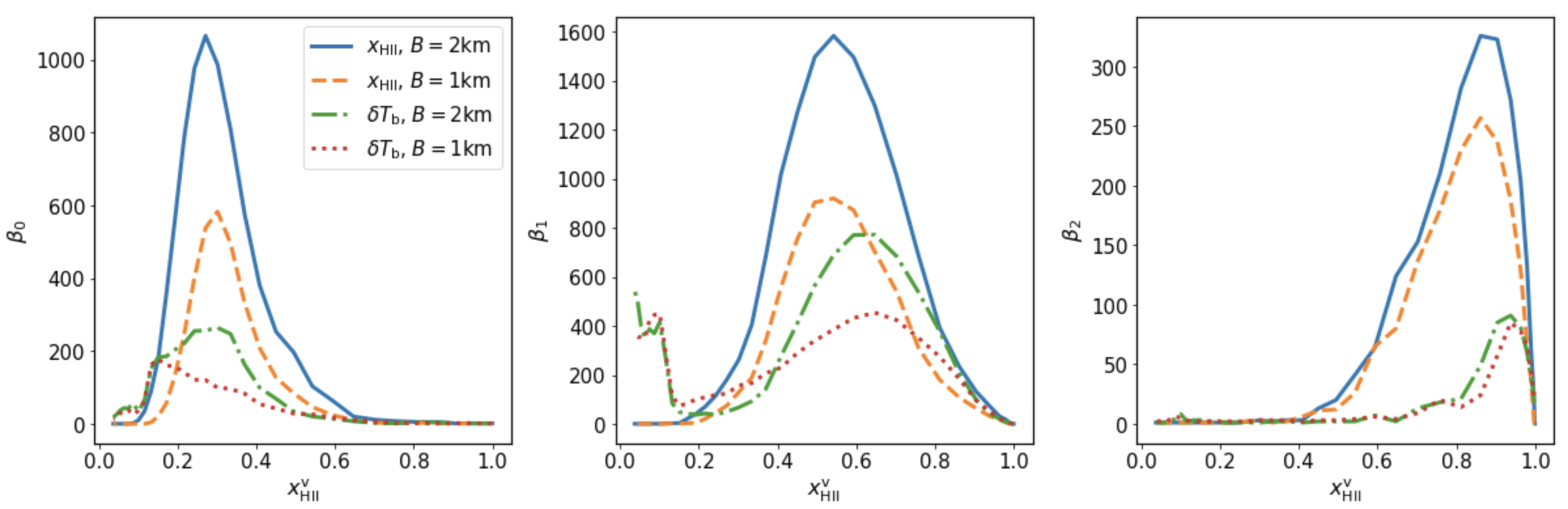}
  \caption{The solid and dashed (dash-dotted and dotted) represent Betti numbers curve estimated from the $x_\mathrm{HII}$ ($\delta T_\mathrm{b}$) fields from simulation FN1 at resolutions corresponding to a maximum baseline $B$ of 2 km and 1 km respectively.}
  \label{fig:betti_dT_LimRes_noNoise}
\end{figure*}

Fig.~\ref{fig:betti_dT_LimRes_noNoise} shows the impact of SKA-Low resolution on the estimated Betti numbers. The solid and dashed lines represent the Betti numbers calculated from the $x_\mathrm{HII}$ data sets degraded to a resolution corresponding to a maximum baseline ($B$) of 2 km and 1 km respectively. These curves differ slightly from the ones shown in Fig.~\ref{fig:betti_xHII_vary_res} as the resolution corresponding to a certain value of $B$ is redshift dependent.
Even though the peak amplitude decreases as the resolution is degraded, the peak position of all the Betti number curves estimated from $x_\mathrm{HII}$ does not change. This behaviour agrees with our findings in Section~\ref{sec:dependence_SimRes}.

The dash-dotted and dotted lines in Fig.~\ref{fig:betti_dT_LimRes_noNoise} represent the Betti numbers estimated from the $\delta T_\mathrm{b}$ fields at a resolution corresponding to $B$ = 2 km and 1 km respectively. The peak position of $\beta_0$ for the $\delta T_\mathrm{b}$ field observed at $B=2~\mathrm{km}$ matches the peak positions of the $\beta_0$ curves derived from the $x_\mathrm{HII}$ fields. However, for the lower resolution case of $B=1~\mathrm{km}$ the $\beta_0$ derived from the $\delta T_\mathrm{b}$ fields attains its peak at a much earlier epoch. This is because the ionized regions are not very large at these early times and for the lower resolution the superpixel method struggles to separate the ionization and density fluctuations. This effect is in fact also present for the $B=2~\mathrm{km}$ case where it causes the non-smooth increase of \bettiZero.

The middle panel of Fig.~\ref{fig:betti_dT_LimRes_noNoise} shows that the $\beta_1$ curves for $\delta T_\mathrm{b}$ field are slightly biased towards larger $x^\mathrm{v}_\mathrm{HII}$ compared to the ones for $x_\mathrm{HII}$ field. We also observe that the peak amplitude of $\beta_1$ curves for $\delta T_\mathrm{b}$ field decreases and the width of the curve increases compared to the ones for $x_\mathrm{HII}$ field. At early times ($x^\mathrm{v}_\mathrm{HII}\lesssim 0.2$), the $\beta_1$ determined from $x_\mathrm{HII}$ field is close to zero for all our reionization scenarios (see Section~\ref{sec:source_models_xHII}). However our method gives non-zero values for $\beta_1$ determined from $\delta T_\mathrm{b}$ field. This anomaly is again caused as the superpixel method struggles to separate ionization and density fluctuations during these early stages of reionization.

The right-most panel of Fig.~\ref{fig:betti_dT_LimRes_noNoise} shows the $\beta_2$ curves for $\delta T_\mathrm{b}$ field. These curves are also slightly biased towards larger $x^\mathrm{v}_\mathrm{HII}$ compared to the ones for $x_\mathrm{HII}$ field.
Even though the resolution corresponding to $B=1~\mathrm{km}$ smooths away more features compared to $B=2~\mathrm{km}$, the superpixel method identifies similar number of neutral regions as these neutral regions are relatively large structures \citep[][]{Giri2019NeutralTomography}. Therefore the $\beta_2$ curves for both resolution cases overlap with each other.

\subsection{Telescope noise}
\label{sec:tele_noise}

\begin{figure*}
  \centering
  \includegraphics[width=0.98\textwidth]{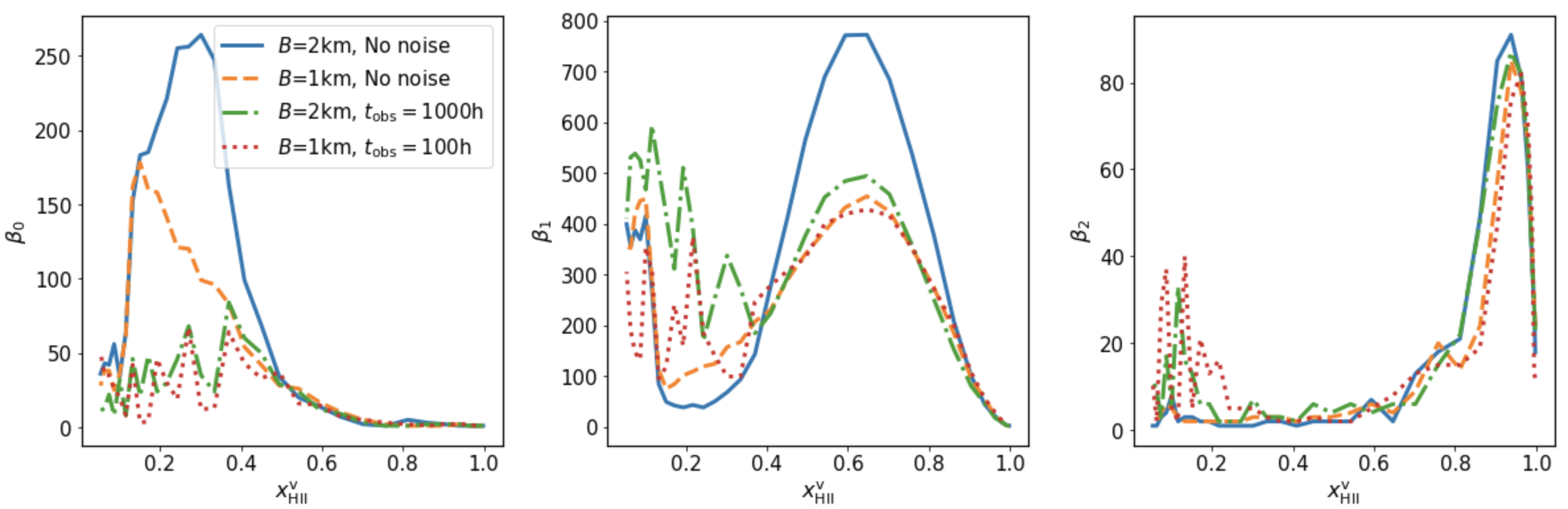}
  \caption{Betti numbers estimated from the $\delta T_\mathrm{b}$ field from simulation FN1 when observed with SKA-Low. The solid and dashed lines represent Betti numbers estimated from noiseless $\delta T_\mathrm{b}$ observed at resolution corresponding to a maximum SKA-Low baseline of 2 km and 1 km respectively. The dash-dotted and dotted lines represent Betti numbers estimated from 1000 h and 100 h observation of $\delta T_\mathrm{b}$ observed at resolution corresponding to a maximum SKA-Low baseline $B$ of 2 km and 1 km respectively.}
  \label{fig:betti_dT_LimRes_noisy}
\end{figure*}

The SKA-Low observations will contain instrumental noise, which we model using the method described in Section~\ref{sec:noise_model}.
Here we study the Betti numbers estimated from noisy 21-cm images. We consider two observation times, which are $t_\mathrm{obs}=1000$ h and 100 h. The 21-cm images observed at $t_\mathrm{obs}=1000$ h is degraded to a resolution corresponding to $B=2$ km. 
In order to get similar SNR, we use $B=1$ km for the 21-cm images observed with $t_\mathrm{obs}=100$ h.

In the left-most, middle and right-most panels of Fig.~\ref{fig:betti_dT_LimRes_noisy}, we show from left to right the $\beta_0$, $\beta_1$ and $\beta_2$ curves estimated from $\delta T_\mathrm{b}$ field. As can be seen from these curves our method can reconstruct the topology of reionization most reliably for epochs $x^\mathrm{v}_\mathrm{HII}\gtrsim 0.4$ from the noisy 21-cm images. For earlier times \RefReport{we find noisy Betti curves, which is what we expect for data sets in which the noise dominates over the signal.} In Section~\ref{sec:betti_xHII}, we found that the topology of reionization during the early stages is mostly traced by the $\beta_0$ values. These are very difficult to extract from the noisy 21-cm images as most ionized regions are small scale features which are indistinguishable from noise. 
The $\beta_1$ and $\beta_2$ values are more sensitive to the topology of reionization during the middle and late stages of reionization respectively. We observe good reconstruction of $\beta_1$ and $\beta_2$ curves in Fig.~\ref{fig:betti_dT_LimRes_noisy} derived from noisy 21-cm images, with the exception of the \bettiOne values for the $B=2$~km, 1000~h observation which are severely underestimated. 

As the $\beta_0$ curves at $x^\mathrm{v}_\mathrm{HII} \lesssim 0.4$ are noisy, we fit Eq.~\ref{eq:fit_b0} to the these curves for $x^\mathrm{v}_\mathrm{HII} \gtrsim 0.4$ only. We list the fit parameter values in Table~\ref{tab:fitted_params}.
Even though we cannot see any clear peak position in the $\beta_0$ curves, the fit for the mock observations $t_\mathrm{obs} = 1000$ h does give us good values for $\mu_0$ and $\sigma_0$. \RefReport{Even though the curve derived for $t_\mathrm{obs} = 100$ h seems to resemble the one for $t_\mathrm{obs} = 1000$ h, our cross-validation technique shows that no reasonable fit can be made}. We therefore conclude that to study the $\beta_0$ evolution from 21-cm images, a long observation time is needed.

Similarly, we fitted parameters of Eqs.~\ref{eq:fit_b1} and \ref{eq:fit_b2} to the $\beta_1$ and $\beta_2$ curves for $x^\mathrm{v}_\mathrm{HII} \gtrsim 0.4$. These fit parameters are also given in Table~\ref{tab:fitted_params}.
The peak position and width of the $\beta_1$ curves derived from the noisy 21-cm images match quite well with the ones for the noiseless case. However, their amplitude is lower than for the ideal case. The $\beta_2$ curves derived from noisy 21-cm images almost overlap with the curves derived from the noiseless cases, leading to very similar fitting parameters for $\beta_2$ for all cases, irrespective of observation time. This indicates that we can study the topology of reionization during its middle and late stages even with 21-cm observations with relatively short observation times.

\subsection{Comparing different reionization models}

\begin{figure*}
  \centering
  \includegraphics[width=0.98\textwidth]{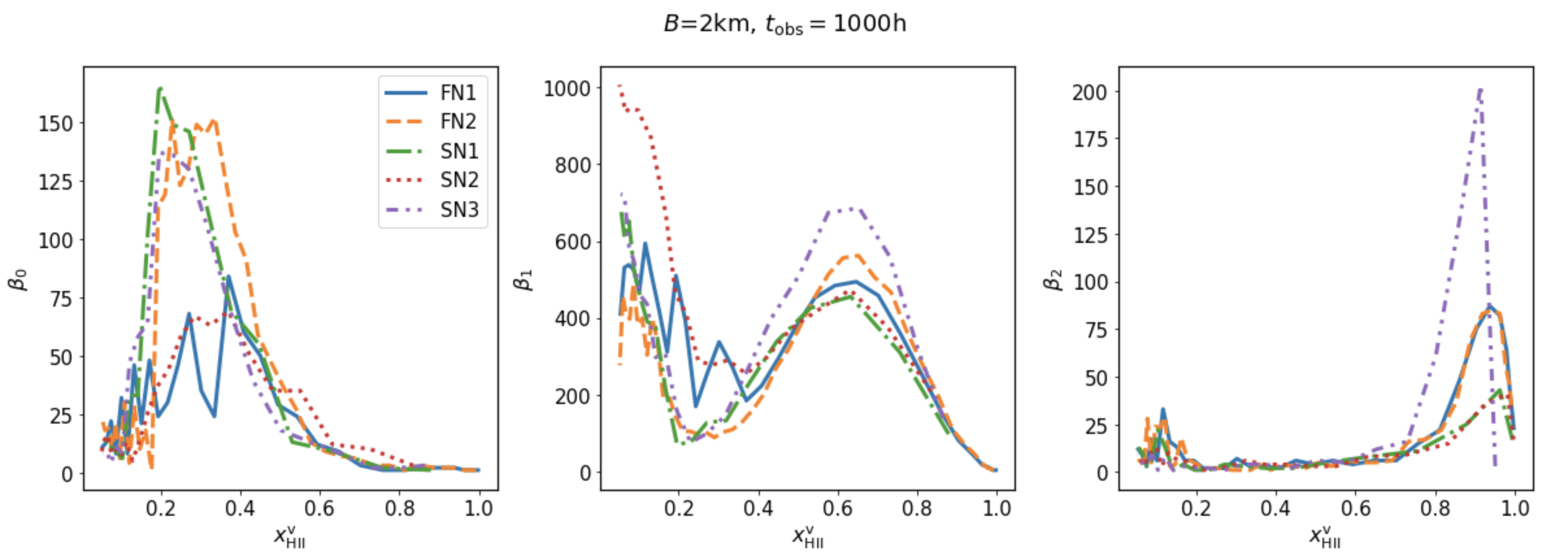}\\
  \includegraphics[width=0.98\textwidth]{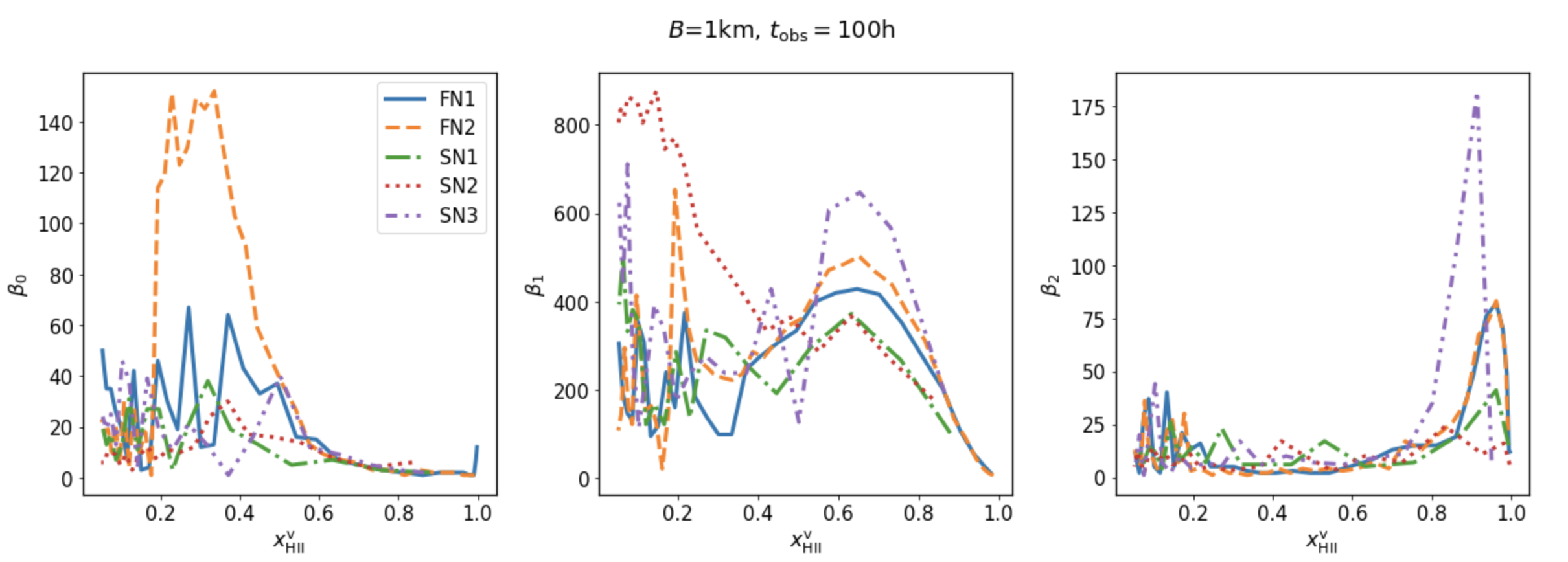}
  \caption{Betti numbers estimated from the $\delta T_\mathrm{b}$ field for all our reionization models when observed with SKA-Low. The top and bottom panels represent the Betti number curves extracted from mock observations of $t_\mathrm{obs} = 1000$ h ($B=2$ km) and $t_\mathrm{obs} = 100$ h ($B=1$ km) respectively. Different models are marked with different line-styles that are shown in the legend.}
  \label{fig:betti_dT_LimRes_noisy_source_models}
\end{figure*}

In this section, we compare Betti number curves for all our reionization models when observed with SKA-Low. The top panel of Fig.~\ref{fig:betti_dT_LimRes_noisy_source_models} shows the curves derived from mock observations at $t_\mathrm{obs} = 1000$ h and $B=2$ km. We find that we can extract the correct evolution the topology of reionization in all the models. These curves are more reliable at epochs $x^\mathrm{v}_\mathrm{HII}\gtrsim 0.4$. Using all the Betti numbers together, we can distinguish between the models. In Table~\ref{tab:fitted_params}, we give the parameter values while fitting all these curves with Eq.~\ref{eq:fit_b0}, \ref{eq:fit_b1} and \ref{eq:fit_b2}. The derived $\mu$ and $\sigma$ values are quite close to the values derived from the $x_\mathrm{HII}$ field. 

We also compare the Betti number curves for all our reionization models in our more pessimistic scenario. The bottom panel of Fig.~\ref{fig:betti_dT_LimRes_noisy_source_models} gives the curves derived from mock observations at $t_\mathrm{obs} = 100$ h and $B=1$ km. Even though the curves are noisy, we can discern the expected evolution of the Betti numbers for each model. These curves are more reliable at epochs $x^\mathrm{v}_\mathrm{HII}\gtrsim 0.5$, making it difficult to derive a fit for the evolution of $\beta_0$. 
In Table~\ref{tab:fitted_params} contains the fit parameters for the $\beta_1$ and $\beta_2$ curves.
As the late neutral islands tend to be large, we find that we can measure the topology of reionization at late times quite well. As noted above, this can even be done using relatively short observations times with SKA-Low.

\section{Summary}
\label{sec:summary}


In this work we explored a new method to study the three dimensional topology of \HII regions during reionization, namely the three Betti numbers. In the perspective of the distribution of ionized regions the zeroth Betti number (\bettiZero) gives the number of isolated \HII regions, the first Betti number (\bettiOne) gives the number of neutral tunnels through ionized regions and the second Betti number (\bettiTwo) gives the number of neutral islands embedded in ionized regions.

We investigated the evolution of the Betti numbers in a small set of fully numerical and semi-numerical simulations and found a characteristic pattern in which each of the Betti numbers first rises and then falls, each attaining a maximum value at different stages of reionization. When reionization starts, most of the ionised bubbles are isolated and \bettiZero is the largest of the three numbers. It initially rises as the number of ionising sources, including individual and clustered sources, grows. However, as existing ionized regions grow and new ones continue to form, they will start to overlap, leading eventually to a decrease of \bettiZero.

When substantial overlapping of ionised bubbles starts, more and more of these connected regions will be crossed by neutral tunnels and we consequently observe a rise in the values of \bettiOne. The time when the decreasing \bettiZero and the increasing \bettiOne cross corresponds to the appearance of  the ionized percolation cluster \citep[][]{Furlanetto2016ReionizationTheory}. \bettiOne reaches its maximum value at the middle stages of reionization. During this stage, isolated neutral regions, some of them the remnants of broken up tunnels, become more common and therefore the value of the second Betti number (\bettiTwo) starts to increase. The time when the \bettiOne and \bettiTwo curves cross corresponds to the disappearance of the neutral percolation cluster. During the later stages of reionization \bettiTwo reaches its peak value after which it drops as the last neutral islands disappear. 

We find that the characteristic evolution of the Betti numbers with the volume-averaged ionisation fraction (\XHIIv) of the Universe is quite well fitted by simple functions, a normal or Gaussian function for \bettiOne and log-normal functions for \bettiZero and \bettiTwo. Each of these is determined by three parameters specifying the peak amplitude ($A_i$), peak position ($\mu_i$) and width ($\sigma_i$) of the curves. Using our suite of reionization simulations, we showed that we can distinguish reionization models by their sets of Betti curves.

On the basis of these results we conclude that the Betti numbers provide a better way to describe the topology of reionization than the much better studied genus or Euler characteristic. Not only do the Betti numbers represent a set of intuitively clear morphological concepts in the context of ionized regions, they also display characteristic evolutionary patterns in which they each distinguish different stages of reionization, and which also appear to fit simple functional forms. Furthermore, their evolution shows a connection with the percolation process characteristic of reionization.
As the Euler characteristic can easily be calculated from the Betti numbers, there appears to be no advantage to studying the topology of reionization with it.

The redshifted 21-cm signal probes the distribution of neutral hydrogen gas during reionization. Observations with the future SKA-Low will deliver 3-D tomographic data sets consisting of sky images of the 21-cm signal at different frequencies. We explored the prospect of deriving the topology of \HII regions from such data sets. 
To identify ionized regions in the 21-cm data we use the superpixels method introduced in \citet{Giri2018OptimalObservations}. We find that we can retrieve the topology of reionization quite well for middle and late stages of reionization ($x^\mathrm{v}_\mathrm{HII} \gtrsim 0.4$). At early times, most \HII regions are small and difficult to identify in the mock 21-cm data sets due to a combination of low resolution and instrumental noise. We tested our method for two observation times, 1000 h and 100 h. For both cases we could retrieve the evolution of \bettiOne and \bettiTwo better than the one for \bettiZero. We therefore expect that \bettiOne and \bettiTwo in practice will be more robust measures of topology than \bettiZero.

All Betti numbers curves derived from the mock 21-cm observations are noisy due to the imperfections of the observational data which do not allow a perfect identification of \HII regions. However when we fit \bettiOne and \bettiTwo curves with our three parameter models, the intrinsic values of the model parameters are retrieved quite well. Retrieving the model parameters for \bettiZero is more challenging for the reasons explained above.

We have also compared the topologies determined from mock 21-cm observations constructed from our suite of reionization simulations. We can clearly distinguish between these models for a 1000 h observation with SKA-Low. Determining the evolution of the Betti numbers with a 100 h observation is more difficult. However, we can still distinguish between our models using \bettiOne and \bettiTwo. As isolated neutral regions are relatively large structures \citep[][]{Giri2019NeutralTomography}, the values of \bettiTwo estimated from the mock observations are the least affected by limited resolution and noise. 

In this paper we have explored a small set of five reionization simulations, produced by two different codes. One might wonder about the validity of some of our conclusions, e.g.\ regarding the shapes of the Betti curves. \RefReport{If reionization proceeded inside-out we expect the Betti curves to always show the behaviour we have found: each Betti number peaks at a different time, in the order \bettiZero, \bettiOne, \bettiTwo. The reason for this is that for inside-out reionization there will be a strong correlation between the local density and the redshift of reionization. The areas of highest density reionize first and the ones of lowest density reionize last. This implies that different stages of reionization can approximately be connected to different threshold values in the density field, where everything above that threshold value is ionized. In this case we expect the evolution of the Betti numbers of ionized regions to approximately follow the Betti curves for different threshold values in a GRF shown in Fig.~\ref{fig:betti_of_grf}. This means that  each Betti number will peak at a different time, in the order \bettiZero, \bettiOne, \bettiTwo, just as we have found for all our models. As reionization is a radiative transfer process, the correlation between density and redshift of reionziation will never be perfect, which is why the actual curves are not identical to the more or less Gaussian curves from Fig.~\ref{fig:betti_of_grf}. Why the evolutionary Betti curves would favour log-normal shapes for \bettiZero and \bettiTwo and a Gaussian shape for \bettiOne is harder to explain. We consider our proposed fitting functions to be a useful working hypothesis until cases are found where this approach  breaks down.}

Our study has used a number of simplifying assumptions which we would like to briefly discuss here. The first is that we have calculated the differential brightness temperature assuming the high spin temperature limit. If X-ray heating is not as efficient as generally assumed, this assumption may not be globally valid during the earlier stages of reionization. However, as we have seen, the early stages of reionization will anyway be challenging to study topologically as it will be difficult to identify ionized regions in the data.

We have also neglected the various line-of-sight effects when constructing our mock 21-cm observations. These are the light-cone (LC) effect \citep[e.g.][]{Datta2012Light-coneSpectrum} and redshift space distortions (RSD) \citep[e.g.][]{Jensen2013ProbingDistortions}. 
The former effect is caused by the evolution of the signal along the frequency direction. If we would analyse 21-cm data sets with a large frequency range ($\gtrsim 10$~MHz), we would see more overlapping regions at the high frequency end than at the low frequency end. This will affect the topology of \HII regions and therefore the Betti numbers. However, this can be avoided by choosing a narrower frequency range. The RSD effect displaces the signal from its cosmological redshift due to peculiar velocities of gas. This process mostly alters the amplitude of the signal and has a relatively minor impact on the  morphology of \HII regions \citep[][]{Giri2018BubbleTomography} 
Therefore we expect only a minor impact on the Betti numbers, specially at the resolution of the telescope. We defer a detailed study of the impact of these line-of-sight effects to the future.

A potentially larger challenge is the impact from bright foreground signals which are many times brighter than the 21-cm signal we want to analyse \citep[][]{Jelic2008ForegroundExperiment}. A good foreground mitigation strategy is crucial to extract the 21-cm signal from radio observation. However, these methods are not perfect and therefore will leave residuals in the 21-cm images. If our structure identification methods struggles to distinguish between these residual foreground and the \HII regions, then studying the topology of reionization would become difficult. In such a case, we have to develop a foreground mitigation method that preserves the topological information. A detailed study of the impact of foreground residual is beyond the scope of this paper.

\section*{Acknowledgements}
We acknowledge Rien van de Weygaert for useful discussion. GM is supported in part by Swedish Research Council grant 2016-03581. The computations were enabled by resources provided by the Swedish National Infrastructure for Computing (SNIC) at PDC partially funded by the Swedish Research Council through grant agreement no. 2018-05973. We also have used results obtained using the PRACE Research Infrastructure resources Curie based at the Très Grand Centre de Calcul (TGCC) operated by CEA near Paris, France and Marenostrum based in the Barcelona Supercomputing Center, Spain. Time on these resources was awarded by PRACE under PRACE4LOFAR grants 2012061089 and 2014102339 as well as under the Multi-scale Reionization grants 2014102281 and 2015122822. 

\section*{DATA AVAILABILITY}
The data underlying this article will be shared on reasonable request to the corresponding author.



\bibliographystyle{mnras}
\bibliography{references,mendeley} 

\begin{thebibliography}{}
\makeatletter
\relax
\def\mn@urlcharsother{\let\do\@makeother \do\$\do\&\do\#\do\^\do\_\do\%\do\~}
\def\mn@doi{\begingroup\mn@urlcharsother \@ifnextchar [ {\mn@doi@}
  {\mn@doi@[]}}
\def\mn@doi@[#1]#2{\def\@tempa{#1}\ifx\@tempa\@empty \href
  {http://dx.doi.org/#2} {doi:#2}\else \href {http://dx.doi.org/#2} {#1}\fi
  \endgroup}
\def\mn@eprint#1#2{\mn@eprint@#1:#2::\@nil}
\def\mn@eprint@arXiv#1{\href {http://arxiv.org/abs/#1} {{\tt arXiv:#1}}}
\def\mn@eprint@dblp#1{\href {http://dblp.uni-trier.de/rec/bibtex/#1.xml}
  {dblp:#1}}
\def\mn@eprint@#1:#2:#3:#4\@nil{\def\@tempa {#1}\def\@tempb {#2}\def\@tempc
  {#3}\ifx \@tempc \@empty \let \@tempc \@tempb \let \@tempb \@tempa \fi \ifx
  \@tempb \@empty \def\@tempb {arXiv}\fi \@ifundefined
  {mn@eprint@\@tempb}{\@tempb:\@tempc}{\expandafter \expandafter \csname
  mn@eprint@\@tempb\endcsname \expandafter{\@tempc}}}

\bibitem[\protect\citeauthoryear{Bandyopadhyay, Choudhury  \&
  Seshadri}{Bandyopadhyay et~al.}{2017}]{Bandyopadhyay2017StudyingDimensions}
Bandyopadhyay B.,  Choudhury T.~R.,   Seshadri T.~R.,  2017, \mn@doi [Monthly
  Notices of the Royal Astronomical Society] {10.1093/mnras/stw3347}, 466, 2302

\bibitem[\protect\citeauthoryear{Barkana}{Barkana}{2016}]{Barkana2016TheCosmology}
Barkana R.,  2016, \mn@doi [Physics Reports] {10.1016/j.physrep.2016.06.006},
  645, 1

\bibitem[\protect\citeauthoryear{Barkana}{Barkana}{2018}]{Barkana2018PossibleStars}
Barkana R.,  2018, \mn@doi [Nature] {10.1038/nature25791}, 555, 71

\bibitem[\protect\citeauthoryear{Betti}{Betti}{1870}]{Betti1870SopraDimensioni}
Betti E.,  1870, \mn@doi [Annali di Matematica Pura ed Applicata (1867-1897)]
  {10.1007/BF02420029}, 4, 140

\bibitem[\protect\citeauthoryear{Bobrowski \& Skraba}{Bobrowski \&
  Skraba}{2020}]{Bobrowski2020HomologicalCharacteristic}
Bobrowski O.,  Skraba P.,  2020, \mn@doi [PHYSICAL REVIEW E]
  {10.1103/PhysRevE.101.032304}, 101, 32304

\bibitem[\protect\citeauthoryear{Bond, Cole, Efstathiou  \& Kaiser}{Bond
  et~al.}{1991}]{Bond1991ExcursionFluctuations}
Bond J.~R.,  Cole S.,  Efstathiou G.,   Kaiser N.,  1991, \mn@doi [The
  Astrophysical Journal] {10.1086/170520}, 379, 440

\bibitem[\protect\citeauthoryear{Bowman \& Rogers}{Bowman \&
  Rogers}{2010}]{bowman2010lower}
Bowman J.~D.,  Rogers A. E.~E.,  2010, Nature, 468, 796

\bibitem[\protect\citeauthoryear{Bowman, Rogers, Monsalve, Mozdzen  \&
  Mahesh}{Bowman et~al.}{2018}]{Bowman2018AnSpectrum}
Bowman J.~D.,  Rogers E.~E.,  Monsalve R.~A.,  Mozdzen T.~J.,   Mahesh N.,
  2018, \mn@doi [Nature Publishing Group] {10.1038/nature25792}, 555

\bibitem[\protect\citeauthoryear{Burns et~al.,}{Burns
  et~al.}{2017}]{burns2017space}
Burns J.~O.,  et~al., 2017, The Astrophysical Journal, 844, 33

\bibitem[\protect\citeauthoryear{{Busch}, {Eide}, {Ciardi}  \&
  {Kakiichi}}{{Busch} et~al.}{2020}]{2020MNRAS.498.4533B}
{Busch} P.,  {Eide} M.~B.,  {Ciardi} B.,   {Kakiichi} K.,  2020, \mn@doi
  [\mnras] {10.1093/mnras/staa2599}, \href
  {https://ui.adsabs.harvard.edu/abs/2020MNRAS.498.4533B} {498, 4533}

\bibitem[\protect\citeauthoryear{Chen, Xu, Wang  \& Chen}{Chen
  et~al.}{2019}]{chen2019stages}
Chen Z.,  Xu Y.,  Wang Y.,   Chen X.,  2019, The Astrophysical Journal, 885, 23

\bibitem[\protect\citeauthoryear{{Chingangbam}, {Park}, {Yogendran}  \& {van de
  Weygaert}}{{Chingangbam} et~al.}{2012}]{Chingangbam2012HotandColdSpotCounts}
{Chingangbam} P.,  {Park} C.,  {Yogendran} K.~P.,   {van de Weygaert} R.,
  2012, \mn@doi [\apj] {10.1088/0004-637X/755/2/122}, \href
  {https://ui.adsabs.harvard.edu/abs/2012ApJ...755..122C} {755, 122}

\bibitem[\protect\citeauthoryear{Datta, Mellema, Mao, Iliev, Shapiro  \&
  Ahn}{Datta et~al.}{2012}]{Datta2012Light-coneSpectrum}
Datta K.~K.,  Mellema G.,  Mao Y.,  Iliev I.~T.,  Shapiro P.~R.,   Ahn K.,
  2012, \mn@doi [Monthly Notices of the Royal Astronomical Society]
  {10.1111/j.1365-2966.2012.21293.x}, 424, 1877

\bibitem[\protect\citeauthoryear{Davis, Efstathiou, Frenk  \& White}{Davis
  et~al.}{1985}]{Davis1985TheMatter}
Davis M.,  Efstathiou G.,  Frenk C.~S.,   White S. D.~M.,  1985, \mn@doi [The
  Astrophysical Journal] {10.1086/163168}, 292, 371

\bibitem[\protect\citeauthoryear{DeBoer et~al.,}{DeBoer
  et~al.}{2017}]{deboer2017hydrogen}
DeBoer D.~R.,  et~al., 2017, Publications of the Astronomical Society of the
  Pacific, 129, 45001

\bibitem[\protect\citeauthoryear{Dixon, Iliev, Mellema, Ahn  \& Shapiro}{Dixon
  et~al.}{2016}]{Dixon2016TheReionization}
Dixon K.~L.,  Iliev I.~T.,  Mellema G.,  Ahn K.,   Shapiro P.~R.,  2016,
  \mn@doi [Monthly Notices of the Royal Astronomical Society]
  {10.1093/mnras/stv2887}, 456, 3011

\bibitem[\protect\citeauthoryear{Doroshkevich}{Doroshkevich}{1970}]{doroshkevich1970spatial}
Doroshkevich A.~G.,  1970, Astrophysics, 6, 320

\bibitem[\protect\citeauthoryear{Edelsbrunner \& Harer}{Edelsbrunner \&
  Harer}{2008}]{edelsbrunner2008persistent}
Edelsbrunner H.,  Harer J.,  2008, Contemporary mathematics, 453, 257

\bibitem[\protect\citeauthoryear{Edelsbrunner \& Harer}{Edelsbrunner \&
  Harer}{2010}]{edelsbrunner2010computational}
Edelsbrunner H.,  Harer J.,  2010, Computational topology: an introduction.
American Mathematical Soc.

\bibitem[\protect\citeauthoryear{Elbers \& Van De~Weygaert}{Elbers \& Van
  De~Weygaert}{2019}]{Elbers2019PersistentPhenomenology}
Elbers W.,  Van De~Weygaert R.,  2019, \mn@doi [Monthly Notices of the Royal
  Astronomical Society] {10.1093/mnras/stz908}, 486, 1523

\bibitem[\protect\citeauthoryear{Feng \& Holder}{Feng \&
  Holder}{2018}]{2018ApJ...858L..17F}
Feng C.,  Holder G.,  2018, \mn@doi [{\textbackslash}apjl]
  {10.3847/2041-8213/aac0fe}, 858, L17

\bibitem[\protect\citeauthoryear{Fialkov \& Barkana}{Fialkov \&
  Barkana}{2019}]{2019MNRAS.486.1763F}
Fialkov A.,  Barkana R.,  2019, \mn@doi [{\textbackslash}mnras]
  {10.1093/mnras/stz873}, 486, 1763

\bibitem[\protect\citeauthoryear{Fialkov, Barkana  \& Cohen}{Fialkov
  et~al.}{2018}]{2018PhRvL.121a1101F}
Fialkov A.,  Barkana R.,   Cohen A.,  2018, \mn@doi [Physical Review Letters]
  {10.1103/PhysRevLett.121.011101}, 121, 11101

\bibitem[\protect\citeauthoryear{Fiorio \& Gustedt}{Fiorio \&
  Gustedt}{1996}]{fiorio1996two}
Fiorio C.,  Gustedt J.,  1996, Theoretical Computer Science, 154, 165

\bibitem[\protect\citeauthoryear{Friedrich, Mellema, Alvarez, Shapiro  \&
  Iliev}{Friedrich et~al.}{2011}]{Friedrich2011TopologyReionization}
Friedrich M.~M.,  Mellema G.,  Alvarez M.~A.,  Shapiro P.~R.,   Iliev I.~T.,
  2011, \mn@doi [Monthly Notices of the Royal Astronomical Society]
  {10.1111/j.1365-2966.2011.18219.x}, 413, 1353

\bibitem[\protect\citeauthoryear{Furlanetto \& Oh}{Furlanetto \&
  Oh}{2016}]{Furlanetto2016ReionizationTheory}
Furlanetto S.~R.,  Oh S.~P.,  2016, \mn@doi [Monthly Notices of the Royal
  Astronomical Society] {10.1093/mnras/stw104}, 457, 1813

\bibitem[\protect\citeauthoryear{Furlanetto, Zaldarriaga  \&
  Hernquist}{Furlanetto et~al.}{2004}]{2004ApJ...613....1F}
Furlanetto S.~R.,  Zaldarriaga M.,   Hernquist L.,  2004, \mn@doi [The
  Astronomical Journal] {10.1086/423025}, 613, 1

\bibitem[\protect\citeauthoryear{Furlanetto, Oh  \& Briggs}{Furlanetto
  et~al.}{2006}]{Furlanetto2006CosmologyUniverse}
Furlanetto S.~R.,  Oh S.~P.,   Briggs F.~H.,  2006, \mn@doi [Physics Reports]
  {10.1016/j.physrep.2006.08.002}, 433, 181

\bibitem[\protect\citeauthoryear{Ghara, Choudhury, Datta  \& Choudhuri}{Ghara
  et~al.}{2017}]{Ghara2017ImagingSKA}
Ghara R.,  Choudhury T.~R.,  Datta K.~K.,   Choudhuri S.,  2017, \mn@doi
  [Monthly Notices of the Royal Astronomical Society] {10.1093/mnras/stw2494},
  464, 2234

\bibitem[\protect\citeauthoryear{Ghara et~al.,}{Ghara
  et~al.}{2020}]{Ghara2020ConstrainingObservations}
Ghara R.,  et~al., 2020, \mn@doi [Monthly Notices of the Royal Astronomical
  Society] {10.1093/mnras/staa487}, 493, 4728

\bibitem[\protect\citeauthoryear{{Ghara}, {Giri}, {Ciardi}, {Mellema}  \&
  {Zaroubi}}{{Ghara} et~al.}{2021}]{2021arXiv210307483G}
{Ghara} R.,  {Giri} S.~K.,  {Ciardi} B.,  {Mellema} G.,   {Zaroubi} S.,  2021,
  arXiv e-prints, \href {https://ui.adsabs.harvard.edu/abs/2021arXiv210307483G}
  {p. arXiv:2103.07483}

\bibitem[\protect\citeauthoryear{Giri}{Giri}{2019}]{giri2019tomographic}
Giri S.~K.,  2019, PhD thesis, Department of Astronomy, Stockholm University

\bibitem[\protect\citeauthoryear{Giri, Mellema, Dixon  \& Iliev}{Giri
  et~al.}{2018a}]{Giri2018BubbleTomography}
Giri S.~K.,  Mellema G.,  Dixon K.~L.,   Iliev I.~T.,  2018a, \mn@doi [MNRAS]
  {10.1093/mnras/stx2539}, 473, 2949

\bibitem[\protect\citeauthoryear{Giri, Mellema  \& Ghara}{Giri
  et~al.}{2018b}]{Giri2018OptimalObservations}
Giri S.~K.,  Mellema G.,   Ghara R.,  2018b, \mn@doi [Monthly Notices of the
  Royal Astronomical Society] {10.1093/mnras/sty1786}, 479, 5596

\bibitem[\protect\citeauthoryear{Giri, Mellema, Aldheimer, Dixon  \&
  Iliev}{Giri et~al.}{2019a}]{Giri2019NeutralTomography}
Giri S.~K.,  Mellema G.,  Aldheimer T.,  Dixon K.~L.,   Iliev I.~T.,  2019a,
  \mn@doi [Monthly Notices of the Royal Astronomical Society]
  {10.1093/mnras/stz2224}, 489, 1590

\bibitem[\protect\citeauthoryear{Giri, D'Aloisio, Mellema, Komatsu, Ghara  \&
  Majumdar}{Giri et~al.}{2019b}]{Giri2019Position-dependentReionization}
Giri S.~K.,  D'Aloisio A.,  Mellema G.,  Komatsu E.,  Ghara R.,   Majumdar S.,
  2019b, \mn@doi [Journal of Cosmology and Astroparticle Physics]
  {10.1088/1475-7516/2019/02/058}, 2019, 058

\bibitem[\protect\citeauthoryear{Giri, Mellema  \& Jensen}{Giri
  et~al.}{2020a}]{Giri2020tools21cm}
Giri S.~K.,  Mellema G.,   Jensen H.,  2020a, \mn@doi [Journal of Open Source
  Software] {10.21105/joss.02363}, 5, 2363

\bibitem[\protect\citeauthoryear{{Giri}, {Zackrisson}, {Binggeli}, {Pelckmans}
  \& {Cubo}}{{Giri} et~al.}{2020b}]{2020MNRAS.491.5277G}
{Giri} S.~K.,  {Zackrisson} E.,  {Binggeli} C.,  {Pelckmans} K.,   {Cubo} R.,
  2020b, \mn@doi [\mnras] {10.1093/mnras/stz3416}, \href
  {https://ui.adsabs.harvard.edu/abs/2020MNRAS.491.5277G} {491, 5277}

\bibitem[\protect\citeauthoryear{Gleser, Nusser, Ciardi  \& Desjacques}{Gleser
  et~al.}{2006}]{Gleser2006TheFunctionals}
Gleser L.,  Nusser A.,  Ciardi B.,   Desjacques V.,  2006, \mn@doi [Mon. Not.
  R. Astron. Soc] {10.1111/j.1365-2966.2006.10556.x}, 370, 1329

\bibitem[\protect\citeauthoryear{{Gnedin}}{{Gnedin}}{2000}]{gnedin2000stagesEoR}
{Gnedin} N.~Y.,  2000, \mn@doi [\apj] {10.1086/308876}, \href
  {https://ui.adsabs.harvard.edu/abs/2000ApJ...535..530G} {535, 530}

\bibitem[\protect\citeauthoryear{Gonzalez-Lorenzo, Juda, Bac, Mari  \&
  Real}{Gonzalez-Lorenzo et~al.}{2016}]{Gonzalez-Lorenzo2016FastComplexes}
Gonzalez-Lorenzo A.,  Juda M.,  Bac A.,  Mari J.~L.,   Real P.,  2016, in
  Lecture Notes in Computer Science (including subseries Lecture Notes in
  Artificial Intelligence and Lecture Notes in Bioinformatics). pp 130--139,
  \mn@doi{10.1007/978-3-319-39441-1{\_}12}, \url
  {https://hal-amu.archives-ouvertes.fr/hal-01341014/document}

\bibitem[\protect\citeauthoryear{Gott, Dickinson  \& Melott}{Gott
  et~al.}{1986}]{Gott1986TheUniverse}
Gott J.~R. I.,  Dickinson M.,   Melott A.~L.,  1986, \mn@doi [The Astrophysical
  Journal] {10.1086/164347}, 306, 341

\bibitem[\protect\citeauthoryear{{Gott}, {Mao}, {Park}  \& {Lahav}}{{Gott}
  et~al.}{1992}]{Gott1992topology}
{Gott} J.~R. I.,  {Mao} S.,  {Park} C.,   {Lahav} O.,  1992, \mn@doi [\apj]
  {10.1086/170912}, \href
  {https://ui.adsabs.harvard.edu/abs/1992ApJ...385...26G} {385, 26}

\bibitem[\protect\citeauthoryear{Greig \& Mesinger}{Greig \&
  Mesinger}{2015}]{Greig201521CMMC:Signal}
Greig B.,  Mesinger A.,  2015, \mn@doi [Monthly Notices of the Royal
  Astronomical Society] {10.1093/mnras/stv571}, 449, 4246

\bibitem[\protect\citeauthoryear{Greig, Trott, Barry, Mutch, Pindor, Webster,
  Stuart  \& Wyithe}{Greig et~al.}{2020b}]{Greig2020ExploringSignal}
Greig B.,  Trott C.~M.,  Barry N.,  Mutch S.~J.,  Pindor B.,  Webster R.~L.,
  Stuart J.,   Wyithe B.,  2020b, arXiv:2008.02639

\bibitem[\protect\citeauthoryear{Greig et~al.,}{Greig
  et~al.}{2020a}]{Greig2020InterpretingObservations}
Greig B.,  et~al., 2020a, arXiv:2006.03203, 000, 0

\bibitem[\protect\citeauthoryear{Harnois-D{\'{e}}raps, Pen, Iliev, Merz,
  Emberson  \& Desjacques}{Harnois-D{\'{e}}raps
  et~al.}{2013}]{2013MNRAS.436..540H}
Harnois-D{\'{e}}raps J.,  Pen U.-L.,  Iliev I.~T.,  Merz H.,  Emberson J.~D.,
  Desjacques V.,  2013, \mn@doi [{\textbackslash}mnras]
  {10.1093/mnras/stt1591}, 436, 540

\bibitem[\protect\citeauthoryear{Hastie, Tibshirani  \& Friedman}{Hastie
  et~al.}{2009}]{hastie2009elements}
Hastie T.,  Tibshirani R.,   Friedman J.,  2009, The elements of statistical
  learning: data mining, inference, and prediction.
Springer Science \& Business Media

\bibitem[\protect\citeauthoryear{Hatcher}{Hatcher}{2002}]{hatcher2002algebraic}
Hatcher A.,  2002, Algebraic Topology.
Cambridge University Press

\bibitem[\protect\citeauthoryear{Hills, Kulkarni, Meerburg  \& Puchwein}{Hills
  et~al.}{2018}]{Hills2018ConcernsData}
Hills R.,  Kulkarni G.,  Meerburg P.~D.,   Puchwein E.,  2018, \mn@doi [Nature]
  {10.1038/s41586-018-0796-5}, 564, E32

\bibitem[\protect\citeauthoryear{Hoshen \& Kopelman}{Hoshen \&
  Kopelman}{1976}]{Hoshen1976PercolationAlgorithm}
Hoshen J.,  Kopelman R.,  1976, Physical Review B, 14, 3438

\bibitem[\protect\citeauthoryear{{Hutter}, {Dayal}, {Yepes}, {Gottl{\"o}ber},
  {Legrand}  \& {Ucci}}{{Hutter} et~al.}{2021}]{2021MNRAS.tmp..610H}
{Hutter} A.,  {Dayal} P.,  {Yepes} G.,  {Gottl{\"o}ber} S.,  {Legrand} L.,
  {Ucci} G.,  2021, \mn@doi [\mnras] {10.1093/mnras/stab602}, \href
  {https://ui.adsabs.harvard.edu/abs/2021MNRAS.tmp..610H} {}

\bibitem[\protect\citeauthoryear{Iliev, Mellema, Pen, Merz, Shapiro  \&
  Alvarez}{Iliev et~al.}{2006}]{Iliev2006SimulatingReionization}
Iliev I.~T.,  Mellema G.,  Pen U.~L.,  Merz H.,  Shapiro P.~R.,   Alvarez
  M.~A.,  2006, \mn@doi [Monthly Notices of the Royal Astronomical Society]
  {10.1111/j.1365-2966.2006.10502.x}, 369, 1625

\bibitem[\protect\citeauthoryear{Iliev, Mellema, Ahn, Shapiro, Mao  \&
  Pen}{Iliev et~al.}{2014}]{Iliev2014SimulatingEnough}
Iliev I.~T.,  Mellema G.,  Ahn K.,  Shapiro P.~R.,  Mao Y.,   Pen U.~L.,  2014,
  \mn@doi [Monthly Notices of the Royal Astronomical Society]
  {10.1093/mnras/stt2497}, 439, 725

\bibitem[\protect\citeauthoryear{Jelic et~al.,}{Jelic
  et~al.}{2008}]{Jelic2008ForegroundExperiment}
Jelic V.,  et~al., 2008, \mn@doi [Monthly Notices of the Royal Astronomical
  Society, Volume 389, Issue 3, pp. 1319-1335.]
  {10.1111/j.1365-2966.2008.13634.x}, 389, 1319

\bibitem[\protect\citeauthoryear{Jensen et~al.,}{Jensen
  et~al.}{2013}]{Jensen2013ProbingDistortions}
Jensen H.,  et~al., 2013, \mn@doi [Monthly Notices of the Royal Astronomical
  Society] {10.1093/mnras/stt1341}, 435, 460

\bibitem[\protect\citeauthoryear{Kaczynski, Mischaikow  \& Mrozek}{Kaczynski
  et~al.}{2004}]{kaczynski2004computational}
Kaczynski T.,  Mischaikow K.,   Mrozek M.,  2004, Springer-Verlag, New York, 3,
  35

\bibitem[\protect\citeauthoryear{Kakiichi et~al.,}{Kakiichi
  et~al.}{2017}]{Kakiichi2017RecoveringTomography}
Kakiichi K.,  et~al., 2017, \mn@doi [Monthly Notices of the Royal Astronomical
  Society] {10.1093/mnras/stx1568}, 471, 1936

\bibitem[\protect\citeauthoryear{{Kapahtia}, {Chingangbam}, {Appleby}  \&
  {Park}}{{Kapahtia} et~al.}{2018}]{2018JCAP...10..011K}
{Kapahtia} A.,  {Chingangbam} P.,  {Appleby} S.,   {Park} C.,  2018, \mn@doi
  [\jcap] {10.1088/1475-7516/2018/10/011}, \href
  {https://ui.adsabs.harvard.edu/abs/2018JCAP...10..011K} {2018, 011}

\bibitem[\protect\citeauthoryear{{Kapahtia}, {Chingangbam}  \&
  {Appleby}}{{Kapahtia} et~al.}{2019}]{Kapahtia2019Morphology}
{Kapahtia} A.,  {Chingangbam} P.,   {Appleby} S.,  2019, \mn@doi [\jcap]
  {10.1088/1475-7516/2019/09/053}, \href
  {https://ui.adsabs.harvard.edu/abs/2019JCAP...09..053K} {2019, 053}

\bibitem[\protect\citeauthoryear{{Kapahtia}, {Chingangbam}, {Ghara}, {Appleby}
  \& {Choudhury}}{{Kapahtia} et~al.}{2021}]{2021arXiv210103962K}
{Kapahtia} A.,  {Chingangbam} P.,  {Ghara} R.,  {Appleby} S.,   {Choudhury}
  T.~R.,  2021, arXiv e-prints, \href
  {https://ui.adsabs.harvard.edu/abs/2021arXiv210103962K} {p. arXiv:2101.03962}

\bibitem[\protect\citeauthoryear{Keating, Weinberger, Kulkarni, Haehnelt,
  Chardin  \& Aubert}{Keating et~al.}{2019}]{Keating2019LongHydrogen}
Keating L.~C.,  Weinberger L.~H.,  Kulkarni G.,  Haehnelt M.~G.,  Chardin J.,
  Aubert D.,  2019, arxiv:1905.12640

\bibitem[\protect\citeauthoryear{Komatsu et~al.,}{Komatsu
  et~al.}{2011}]{2011ApJS..192...18K}
Komatsu E.,  et~al., 2011, \mn@doi [{\textbackslash}apjs]
  {10.1088/0067-0049/192/2/18}, 192, 18

\bibitem[\protect\citeauthoryear{Koopmans et~al.,}{Koopmans
  et~al.}{2015}]{Koopmans2015TheArray}
Koopmans L.,  et~al., 2015, in Advancing Astrophysics with the Square Kilometre
  Array (AASKA14). \url {http://pos.sissa.it/}

\bibitem[\protect\citeauthoryear{Kulkarni, Keating, Haehnelt, Bosman, Puchwein,
  Chardin  \& Aubert}{Kulkarni et~al.}{2019}]{Kulkarni2019Large5.5}
Kulkarni G.,  Keating L.~C.,  Haehnelt M.~G.,  Bosman S. E.~I.,  Puchwein E.,
  Chardin J.,   Aubert D.,  2019, \mn@doi [Monthly Notices of the Royal
  Astronomical Society: Letters] {10.1093/mnrasl/slz025}, 485, L24

\bibitem[\protect\citeauthoryear{Lee, Cen, Gott~III  \& Trac}{Lee
  et~al.}{2008}]{Lee2008TheReionization}
Lee K.,  Cen R.,  Gott~III J.~R.,   Trac H.,  2008, \mn@doi [The Astrophysical
  Journal] {10.1086/525520}, 675, 8

\bibitem[\protect\citeauthoryear{Lim \& Mellema}{Lim \&
  Mellema}{2003}]{lim20033d}
Lim A.,  Mellema G.,  2003, Astronomy \& Astrophysics, 405, 189

\bibitem[\protect\citeauthoryear{Majumdar, Pritchard, Mondal, Watkinson,
  Bharadwaj  \& Mellema}{Majumdar
  et~al.}{2018}]{Majumdar2018QuantifyingBispectrum}
Majumdar S.,  Pritchard J.~R.,  Mondal R.,  Watkinson C.~A.,  Bharadwaj S.,
  Mellema G.,  2018, \mn@doi [Monthly Notices of the Royal Astronomical
  Society] {10.1093/mnras/sty535}, 476, 4007

\bibitem[\protect\citeauthoryear{Makarenko, Shukurov, Henderson, Rodrigues,
  Bushby  \& Fletcher}{Makarenko
  et~al.}{2018}]{Makarenko2018TopologicalDiagrams}
Makarenko I.,  Shukurov A.,  Henderson R.,  Rodrigues L. F.~S.,  Bushby P.,
  Fletcher A.,  2018, \mn@doi [MNRAS] {10.1093/mnras/stx3337}, 475, 1843

\bibitem[\protect\citeauthoryear{{Matsubara}}{{Matsubara}}{1994}]{Matsubara1994Genus}
{Matsubara} T.,  1994, \mn@doi [\apjl] {10.1086/187570}, \href
  {https://ui.adsabs.harvard.edu/abs/1994ApJ...434L..43M} {434, L43}

\bibitem[\protect\citeauthoryear{Matsubara}{Matsubara}{1996}]{MatsubaraTakahikoYokoyama1996GenusFields}
Matsubara Takahiko;~Yokoyama J.,  1996, \mn@doi [THE ASTROPHYSICAL JOURNAL]
  {10.1086/177257}, 463, 409

\bibitem[\protect\citeauthoryear{Mellema, Iliev, Alvarez  \& Shapiro}{Mellema
  et~al.}{2006a}]{2006NewA...11..374M}
Mellema G.,  Iliev I.~T.,  Alvarez M.~A.,   Shapiro P.~R.,  2006a, \mn@doi
  [{\textbackslash}na] {10.1016/j.newast.2005.09.004}, 11, 374

\bibitem[\protect\citeauthoryear{Mellema, Iliev, Pen  \& Shapiro}{Mellema
  et~al.}{2006b}]{Mellema2006SimulatingSignals}
Mellema G.,  Iliev I.~T.,  Pen U.~L.,   Shapiro P.~R.,  2006b, \mn@doi [Monthly
  Notices of the Royal Astronomical Society]
  {10.1111/j.1365-2966.2006.10919.x}, 372, 679

\bibitem[\protect\citeauthoryear{Mellema et~al.,}{Mellema
  et~al.}{2013}]{Mellema2013ReionizationArray}
Mellema G.,  et~al., 2013, \mn@doi [Experimental Astronomy]
  {10.1007/s10686-013-9334-5}, 36, 235

\bibitem[\protect\citeauthoryear{Mellema, Koopmans, Shukla, Datta, Mesinger  \&
  Majumdar}{Mellema et~al.}{2015}]{Mellema2015HISKA}
Mellema G.,  Koopmans L.,  Shukla H.,  Datta K.~K.,  Mesinger A.,   Majumdar
  S.,  2015, in Advancing Astrophysics with the Square Kilometre Array
  (AASKA14). p.~10, \url {http://pos.sissa.it/}

\bibitem[\protect\citeauthoryear{Mertens et~al.,}{Mertens
  et~al.}{2020}]{Mertens2020ImprovedLOFAR}
Mertens F.~G.,  et~al., 2020, \mn@doi [Monthly Notices of the Royal
  Astronomical Society] {10.1093/mnras/staa327}, 493, 1662

\bibitem[\protect\citeauthoryear{Mesinger, Furlanetto  \& Cen}{Mesinger
  et~al.}{2011}]{Mesinger201121CMFAST:Signal}
Mesinger A.,  Furlanetto S.,   Cen R.,  2011, \mn@doi [Monthly Notices of the
  Royal Astronomical Society] {10.1111/j.1365-2966.2010.17731.x}, 411, 955

\bibitem[\protect\citeauthoryear{{Monaghan}}{{Monaghan}}{1992}]{Monaghan1992sph}
{Monaghan} J.~J.,  1992, \mn@doi [\araa] {10.1146/annurev.aa.30.090192.002551},
  \href {https://ui.adsabs.harvard.edu/abs/1992ARA&A..30..543M} {30, 543}

\bibitem[\protect\citeauthoryear{Mondal et~al.,}{Mondal
  et~al.}{2020}]{Mondal2020TightLOFAR}
Mondal R.,  et~al., 2020, Monthly Notices of the Royal Astronomical Society,
  498, 178

\bibitem[\protect\citeauthoryear{Morales \& Wyithe}{Morales \&
  Wyithe}{2010}]{Morales2010ReionizationFluctuations}
Morales M.~F.,  Wyithe J. S.~B.,  2010, \mn@doi [Annual Review of Astronomy and
  Astrophysics] {10.1146/annurev-astro-081309-130936}, 48, 127

\bibitem[\protect\citeauthoryear{Mu{\~{n}}oz \& Loeb}{Mu{\~{n}}oz \&
  Loeb}{2018}]{2018arXiv180210094M}
Mu{\~{n}}oz J.~B.,  Loeb A.,  2018, arXiv e-prints, p. arXiv:1802.10094

\bibitem[\protect\citeauthoryear{{Nasir} \& {D'Aloisio}}{{Nasir} \&
  {D'Aloisio}}{2020}]{2020MNRAS.494.3080N}
{Nasir} F.,  {D'Aloisio} A.,  2020, \mn@doi [\mnras] {10.1093/mnras/staa894},
  \href {https://ui.adsabs.harvard.edu/abs/2020MNRAS.494.3080N} {494, 3080}

\bibitem[\protect\citeauthoryear{{Park}, {Kim}  \& {Gott}}{{Park}
  et~al.}{2005}]{Park2005topology}
{Park} C.,  {Kim} J.,   {Gott} J.~Richard I.,  2005, \mn@doi [\apj]
  {10.1086/452621}, \href
  {https://ui.adsabs.harvard.edu/abs/2005ApJ...633....1P} {633, 1}

\bibitem[\protect\citeauthoryear{Park et~al.,}{Park
  et~al.}{2013}]{Park2013BettiFields}
Park C.,  et~al., 2013, \mn@doi [Journal of the Korean Astronomical Society]
  {10.5303/JKAS.2013.46.3.125}, 46, 125

\bibitem[\protect\citeauthoryear{{Planck Collaboration} et~al.,}{{Planck
  Collaboration} et~al.}{2016}]{PlanckCollaboration2016PlanckHistory}
{Planck Collaboration} et~al., 2016, preprint

\bibitem[\protect\citeauthoryear{{Planck Collaboration} et~al.,}{{Planck
  Collaboration} et~al.}{2018}]{PlanckCollaboration2018PlanckParameters}
{Planck Collaboration} et~al., 2018, arXiv:1807.06209

\bibitem[\protect\citeauthoryear{Pranav, Edelsbrunner, Van De~Weygaert, Vegter,
  Kerber, Jones  \& Wintraecken}{Pranav et~al.}{2017}]{Pranav2017TheNumbers}
Pranav P.,  Edelsbrunner H.,  Van De~Weygaert R.,  Vegter G.,  Kerber M.,
  Jones B. J.~T.,   Wintraecken M.,  2017, \mn@doi [MNRAS]
  {10.1093/mnras/stw2862}, 465, 4281

\bibitem[\protect\citeauthoryear{Pranav et~al.,}{Pranav
  et~al.}{2019}]{Pranav2019TopologyFunctionals}
Pranav P.,  et~al., 2019, \mn@doi [Monthly Notices of the Royal Astronomical
  Society] {10.1093/mnras/stz541}, 485, 4167

\bibitem[\protect\citeauthoryear{Press \& Davis}{Press \&
  Davis}{1982}]{Press1982HowCatalog}
Press W.~H.,  Davis M.,  1982, \mn@doi [The Astrophysical Journal]
  {10.1086/160183}, 259, 449

\bibitem[\protect\citeauthoryear{Price et~al.,}{Price
  et~al.}{2018}]{price2018design}
Price D.,  et~al., 2018, Monthly Notices of the Royal Astronomical Society,
  478, 4193

\bibitem[\protect\citeauthoryear{Pritchard \& Furlanetto}{Pritchard \&
  Furlanetto}{2007}]{Pritchard200721-cmReionization}
Pritchard J.~R.,  Furlanetto S.~R.,  2007, \mn@doi [Monthly Notices of the
  Royal Astronomical Society] {10.1111/j.1365-2966.2007.11519.x}, 376, 1680

\bibitem[\protect\citeauthoryear{Pritchard \& Loeb}{Pritchard \&
  Loeb}{2012}]{Pritchard201221cmCentury}
Pritchard J.~R.,  Loeb A.,  2012, {21cm cosmology in the 21st century},
  \mn@doi{10.1088/0034-4885/75/8/086901}

\bibitem[\protect\citeauthoryear{Raga, Mellema, Arthur, Binette, Ferruit  \&
  Steffen}{Raga et~al.}{1999}]{raga19993d}
Raga A.,  Mellema G.,  Arthur S.,  Binette L.,  Ferruit P.,   Steffen W.,
  1999, Revista Mexicana de Astronomia y Astrofisica, 35, 123

\bibitem[\protect\citeauthoryear{Rohlfs \& Wilson}{Rohlfs \&
  Wilson}{2013}]{rohlfs2013tools}
Rohlfs K.,  Wilson T.,  2013, {Tools of radio astronomy}.
Springer Science {\&} Business Media

\bibitem[\protect\citeauthoryear{Santos, Raposo, Coutinho-Filho, Copelli, Stam
  \& Douw}{Santos et~al.}{2019}]{PhysRevE.100.032414}
Santos F. A.~N.,  Raposo E.~P.,  Coutinho-Filho M.~D.,  Copelli M.,  Stam
  C.~J.,   Douw L.,  2019, \mn@doi [Phys. Rev. E]
  {10.1103/PhysRevE.100.032414}, 100, 032414

\bibitem[\protect\citeauthoryear{Schmalzing \& Buchert}{Schmalzing \&
  Buchert}{1997}]{Schmalzing1997BeyondStructure}
Schmalzing J.,  Buchert T.,  1997, \mn@doi [The Astrophysical Journal]
  {10.1086/310680}, 482, L1

\bibitem[\protect\citeauthoryear{Scoccimarro}{Scoccimarro}{1998}]{scoccimarro1998transients}
Scoccimarro R.,  1998, Monthly Notices of the Royal Astronomical Society, 299,
  1097

\bibitem[\protect\citeauthoryear{Sheth \& Tormen}{Sheth \&
  Tormen}{1999}]{Sheth1999}
Sheth R.~K.,  Tormen G.,  1999, \mn@doi [{\textbackslash}mnras]
  {10.1046/j.1365-8711.1999.02692.x}, 308, 119

\bibitem[\protect\citeauthoryear{{Singh} \& {Subrahmanyan}}{{Singh} \&
  {Subrahmanyan}}{2019}]{Singh2019EDGES}
{Singh} S.,  {Subrahmanyan} R.,  2019, \mn@doi [\apj]
  {10.3847/1538-4357/ab2879}, \href
  {https://ui.adsabs.harvard.edu/abs/2019ApJ...880...26S} {880, 26}

\bibitem[\protect\citeauthoryear{Singh et~al.,}{Singh et~al.}{2017}]{singh2017}
Singh S.,  et~al., 2017, \mn@doi [{\textbackslash}apj]
  {10.3847/2041-8213/aa831b}, 845, L12

\bibitem[\protect\citeauthoryear{{Songaila} \& {Cowie}}{{Songaila} \&
  {Cowie}}{2010}]{2010ApJ...721.1448S}
{Songaila} A.,  {Cowie} L.~L.,  2010, \mn@doi [\apj]
  {10.1088/0004-637X/721/2/1448}, \href
  {https://ui.adsabs.harvard.edu/abs/2010ApJ...721.1448S} {721, 1448}

\bibitem[\protect\citeauthoryear{Sousbie}{Sousbie}{2011}]{sousbie2011persistentI}
Sousbie T.,  2011, Monthly Notices of the Royal Astronomical Society, 414, 350

\bibitem[\protect\citeauthoryear{Sousbie, Pichon  \& Kawahara}{Sousbie
  et~al.}{2011}]{sousbie2011persistentII}
Sousbie T.,  Pichon C.,   Kawahara H.,  2011, Monthly Notices of the Royal
  Astronomical Society, 414, 384

\bibitem[\protect\citeauthoryear{Sullivan, Iliev  \& Dixon}{Sullivan
  et~al.}{2018}]{Sullivan2018UsingReionization}
Sullivan D.,  Iliev I.~T.,   Dixon K.~L.,  2018, \mn@doi [Monthly Notices of
  the Royal Astronomical Society] {10.1093/mnras/stx2324}, 473, 38

\bibitem[\protect\citeauthoryear{Tashiro, Kadota  \& Silk}{Tashiro
  et~al.}{2014}]{2014PhRvD..90h3522T}
Tashiro H.,  Kadota K.,   Silk J.,  2014, \mn@doi [{\textbackslash}prd]
  {10.1103/PhysRevD.90.083522}, 90, 83522

\bibitem[\protect\citeauthoryear{{Tomita}}{{Tomita}}{1986}]{Tomita1986CurvatureInvariants}
{Tomita} H.,  1986, \mn@doi [Progress of Theoretical Physics]
  {10.1143/PTP.76.952}, \href
  {https://ui.adsabs.harvard.edu/abs/1986PThPh..76..952T} {76, 952}

\bibitem[\protect\citeauthoryear{Trott et~al.,}{Trott
  et~al.}{2020}]{Trott2020DeepObservations}
Trott C.~M.,  et~al., 2020, \mn@doi [Monthly Notices of the Royal Astronomical
  Society] {10.1093/mnras/staa414}, 493, 4711

\bibitem[\protect\citeauthoryear{Wagner, Chen  \& Vu{\c{c}}ini}{Wagner
  et~al.}{2012}]{Wagner2012}
Wagner H.,  Chen C.,   Vu{\c{c}}ini E.,  2012, in Peikert R.,  Hauser H.,  Carr
  H.,   Fuchs R.,  eds, , Topological Methods in Data Analysis and
  Visualization II: Theory, Algorithms, and Applications.
Springer Berlin Heidelberg, Berlin, Heidelberg, pp 91--106,
  \mn@doi{10.1007/978-3-642-23175-9{\_}7}, \url
  {https://doi.org/10.1007/978-3-642-23175-9_7}

\bibitem[\protect\citeauthoryear{Watkinson, Giri, Ross, Dixon, Iliev, Mellema
  \& Pritchard}{Watkinson et~al.}{2019}]{Watkinson2019TheHeating}
Watkinson C.~A.,  Giri S.~K.,  Ross H.~E.,  Dixon K.~L.,  Iliev I.~T.,  Mellema
  G.,   Pritchard J.~R.,  2019, \mn@doi [Monthly Notices of the Royal
  Astronomical Society] {10.1093/mnras/sty2740}, 482, 2653

\bibitem[\protect\citeauthoryear{Watson, Iliev, D’Aloisio, Knebe, Shapiro  \&
  Yepes}{Watson et~al.}{2013}]{Watson2013TheAges}
Watson W.~A.,  Iliev I.~T.,  D’Aloisio A.,  Knebe A.,  Shapiro P.~R.,   Yepes
  G.,  2013, \mn@doi [Monthly Notices of the Royal Astronomical Society]
  {10.1093/mnras/stt791}, 433, 1230

\bibitem[\protect\citeauthoryear{{Wayth} et~al.,}{{Wayth}
  et~al.}{2018}]{Wayth2018mwa}
{Wayth} R.~B.,  et~al., 2018, \mn@doi [\pasa] {10.1017/pasa.2018.37}, \href
  {https://ui.adsabs.harvard.edu/abs/2018PASA...35...33W} {35, 33}

\bibitem[\protect\citeauthoryear{{Wilding}, {Nevenzeel}, {van de Weygaert},
  {Vegter}, {Pranav}, {Jones}, {Efstathiou}  \& {Feldbrugge}}{{Wilding}
  et~al.}{2020}]{2020arXiv201112851W}
{Wilding} G.,  {Nevenzeel} K.,  {van de Weygaert} R.,  {Vegter} G.,  {Pranav}
  P.,  {Jones} B. J.~T.,  {Efstathiou} K.,   {Feldbrugge} J.,  2020, arXiv
  e-prints, \href {https://ui.adsabs.harvard.edu/abs/2020arXiv201112851W} {p.
  arXiv:2011.12851}

\bibitem[\protect\citeauthoryear{Wu, Otoo  \& Shoshani}{Wu
  et~al.}{2005}]{wu2005optimizing}
Wu K.,  Otoo E.,   Shoshani A.,  2005, in Medical Imaging 2005: Image
  Processing. pp 1965--1976

\bibitem[\protect\citeauthoryear{Yoshiura, Shimabukuro, Takahashi  \&
  Matsubara}{Yoshiura et~al.}{2017}]{Yoshiura2017StudyingReionization}
Yoshiura S.,  Shimabukuro H.,  Takahashi K.,   Matsubara T.,  2017, \mn@doi
  [Monthly Notices of the Royal Astronomical Society] {10.1093/mnras/stw2701},
  465, 394

\bibitem[\protect\citeauthoryear{{Zaroubi}}{{Zaroubi}}{2013}]{Zaroubi2013eor}
{Zaroubi} S.,  2013, {The Epoch of Reionization}.
Springer Berlin Heidelberg, p.~45, \mn@doi{10.1007/978-3-642-32362-1_2}

\bibitem[\protect\citeauthoryear{Zomorodian \& Carlsson}{Zomorodian \&
  Carlsson}{2005}]{Zomorodian2005ComputingHomology}
Zomorodian A.,  Carlsson G.,  2005, \mn@doi [Discrete {\&} Computational
  Geometry] {10.1007/s00454-004-1146-y}, 33, 249

\bibitem[\protect\citeauthoryear{van Haarlem et~al.,}{van Haarlem
  et~al.}{2013}]{vanHaarlem2013LOFAR:ARray}
van Haarlem M.~P.,  et~al., 2013, \mn@doi [Astronomy {\&} Astrophysics]
  {10.1051/0004-6361/201220873}, 556, A2

\bibitem[\protect\citeauthoryear{{van de Weygaert}, {Platen}, {Vegter},
  {Eldering}  \& {Kruithof}}{{van de Weygaert} et~al.}{2010}]{vdW2010AlphaWeb}
{van de Weygaert} R.,  {Platen} E.,  {Vegter} G.,  {Eldering} B.,   {Kruithof}
  N.,  2010, in 2010 International Symposium on Voronoi Diagrams in Science and
  Engineering. pp 224--234, \mn@doi{10.1109/ISVD.2010.24}

\bibitem[\protect\citeauthoryear{{van de Weygaert} et~al.,}{{van de Weygaert}
  et~al.}{2011}]{vdW2011AlphaWeb}
{van de Weygaert} R.,  et~al., 2011, {Alpha, Betti and the Megaparsec Universe:
  On the Topology of the Cosmic Web}.
pp 60--101, \mn@doi{10.1007/978-3-642-25249-5_3}

\bibitem[\protect\citeauthoryear{van~der Walt et~al.,}{van~der Walt
  et~al.}{2014}]{scikit-image}
van~der Walt S.,  et~al., 2014, \mn@doi [PeerJ] {10.7717/peerj.453}, 2, e453

\makeatother
\end{thebibliography}




\appendix

\section{Topology of Gaussian random field}
\label{sec:topo_grf}

\begin{figure}
  \centering
  \includegraphics[width=0.4\textwidth]{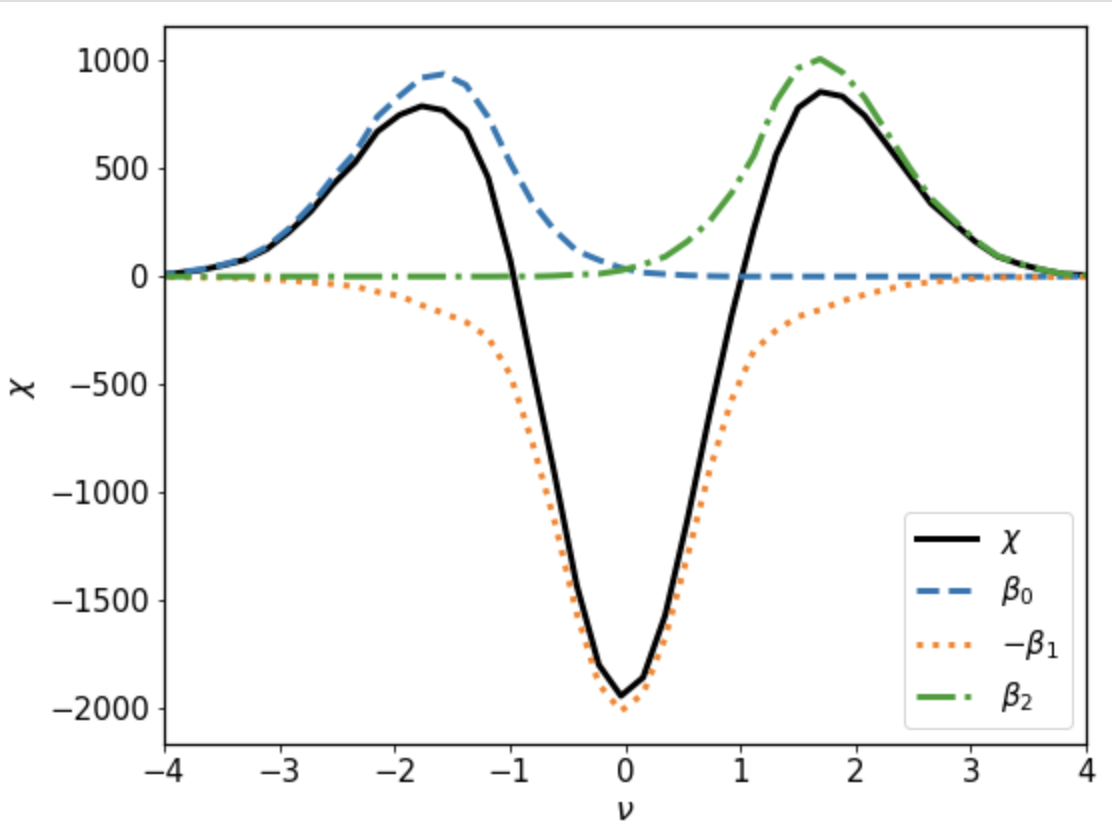}
  \caption{The Euler characteristics $\chi$ (solid, black) and Betti numbers $\beta_0$ (dashed, blue), $\beta_1$ (dotted, red) and $\beta_2$ (dash-dotted, green) of a Gaussian random field calculated for various threshold values $\nu$.}
  \label{fig:betti_of_grf}
\end{figure}

In this section we use a Gaussian random field (GRF) to validate our Betti numbers estimator for digital data. Our GRF is constructed from a $(300)^3$ data cube filled with Gaussian random numbers. Due to discretization, our data cube will not be a perfect GRF. Therefore, following \citet{Park2013BettiFields} we smooth this data with a Gaussian filter with a FWHM of 5 cell widths.


We identify structures by putting a threshold on the intensity values. All the pixels below the threshold constitute the structure whose topology we want to study.
For any field $\mathcal{A}$ the thresholds will scan through all the intensity values. Therefore we will get a set of Betti numbers corresponding to a set of threshold values. The set of thresholds $\nu$ is defined such that $\mathcal{A} = \sqrt{\left \langle \mathcal{A}^2 \right \rangle} \nu + \left \langle \mathcal{A} \right \rangle$. 

The solid curve in Fig.~\ref{fig:betti_of_grf} shows the Euler characteristic ($\chi$) curve of the GRF. The $\chi$ curve is an even function with two peaks and one trough which follows a fixed analytical form \citep[see][]{doroshkevich1970spatial, Tomita1986CurvatureInvariants}. \citet{Giri2019NeutralTomography} already showed that the $\chi$ curve derived by our estimator for digital data matches this analytical form.

Fig.~\ref{fig:betti_of_grf} also shows the Betti numbers $\beta_\kappa$ as a function of threshold  $\nu$ as determined with our estimator (see Section~\ref{sec:topo_estimator}). The dashed, dotted and dash-dotted curve are $\beta_0 (\nu)$, $\beta_1 (\nu)$ and $\beta_2 (\nu)$ respectively. We see that the left peak of the $\chi(\nu)$ curve is caused by the peak in $\beta_0$ or a large number of components compared to tunnels and cavities. Similarly the right peak in $\chi$ is associated with a large value of $\beta_2$ or a large number of cavities compared to components and tunnels. The trough in the $\chi (\nu)$ curve is due to the value of $\beta_1$ or a large number of tunnels compared to components and cavities. The Betti curves for a GRF estimated with our method agree with those found in previous works. As far as we know no analytical forms exist for these $\beta_\kappa(\nu)$ curves. See \citet{Park2013BettiFields} and \citet{Pranav2019TopologyFunctionals} for a more extensive description of the Betti numbers of GRFs.


\bsp	
\label{lastpage}
\end{document}